\pdfoutput=1

\documentclass[sigconf]{acmart}
\AtBeginDocument{%
  \providecommand\BibTeX{{%
    \normalfont B\kern-0.5em{\scshape i\kern-0.25em b}\kern-0.8em\TeX}}}

\copyrightyear{2026}
\acmYear{2026}
\setcopyright{acmcopyright} 
\setcopyright{cc}
\setcctype{by}
\acmConference[CHI '26]{Proceedings of the 2026 CHI Conference on Human Factors in Computing Systems}{April 13--17, 2026}{Barcelona, Spain}
\acmBooktitle{Proceedings of the 2026 CHI Conference on Human Factors in Computing Systems (CHI '26), April 13--17, 2026, Barcelona, Spain}
\acmDOI{10.1145/3772318.3790452}
\acmISBN{979-8-4007-2278-3/2026/04}

\newif\ifrevision
\revisionfalse  

\ifrevision
    \usepackage{xcolor}
    \usepackage{soul}

    \newcommand{\added}[1]{{\color{blue}#1}}

    \newcommand{\deleted}[1]{{\color{red}\st{#1}}}

\else
    \newcommand{\added}[1]{#1}
    \newcommand{\deleted}[1]{}
\fi

\usepackage{graphicx}
\usepackage{diagbox} 
\usepackage{makecell} 
\usepackage{multirow}
\usepackage{colortbl}
\usepackage{makecell}
\usepackage{csquotes}
\usepackage{array}
\usepackage{CJKutf8}
\usepackage{steinmetz}
\usepackage{longtable}
\usepackage{subcaption}
\usepackage{float}
\usepackage{booktabs} 
\usepackage{array}    
\usepackage{multirow} 
\usepackage{tabularx} 
\usepackage{xurl}
\newcommand{\PreserveBackslash}[1]{\let\temp=\\#1\let\\=\temp}

\newcolumntype{C}[1]{>{\PreserveBackslash\centering}p{#1}}
\newcolumntype{R}[1]{>{\PreserveBackslash\raggedleft}p{#1}}
\newcolumntype{L}[1]{>{\PreserveBackslash\raggedright}p{#1}}

\acmSubmissionID{4816}

\begin{document}

\title[Understanding Retired Women's Perceptions of Technology-Enhanced Dance Performance]{From Performers to Creators: Understanding Retired Women's Perceptions of Technology-Enhanced Dance Performance}


\author{Danlin Zheng}
\orcid{0009-0006-8216-1693}
\affiliation{%
  \institution{The Hong Kong University of Science and Technology (Guangzhou)}
  \city{Guangzhou}
  \country{China}
}
\email{dzheng403@connect.hkust-gz.edu.cn}

\author{Xiaoying Wei}
\affiliation{%
  \institution{The Hong Kong University of Science and Technology}
  \city{Hong Kong SAR}
  \country{China}
}
\email{xweias@connect.ust.hk}

\author{Chao Liu}
\affiliation{%
  \institution{The Hong Kong University of Science and Technology (Guangzhou)}
  \city{Guangzhou}
  \country{China}
}
\email{cliu009@connect.hkust-gz.edu.cn}

\author{Quanyu Zhang}
\affiliation{%
  \institution{Xiamen University}
  \city{Xiamen}
  \country{China}
  }
\email{yqiu66666@gmail.com}

\author{Jingling Zhang}
\affiliation{%
  \institution{The Hong Kong University of Science and Technology (Guangzhou)}
  \city{Guangzhou}
  \country{China}
  }
\email{jzhang898@connect.hkust-gz.edu.cn}

\author{Shihui Guo}
\affiliation{%
  \institution{Xiamen University}
  \city{Xiamen}
  \country{China}}
\email{guoshihui@xmu.edu.cn}

\author{Mingming Fan}
\authornote{Corresponding author.}
\affiliation{%
  \institution{The Hong Kong University of Science and Technology (Guangzhou)}
  \city{Guangzhou}
  \country{China}}
\affiliation{%
  \institution{The Hong Kong University of Science and Technology}
  \city{Hong Kong SAR}
  \country{China}
}
\email{mingmingfan@ust.hk}



\begin{abstract}
Over 100 million retired women in China engage in dance, but their performances are constrained by limited resources and age-related decline. While interactive dance technologies can enhance artistic expression, existing systems are largely inaccessible to non-professional older dancers. This paper explores how interactive dance technologies can be designed with an \deleted{aging-friendly}\added{age-sensitive} approach to support retired women in enhancing their stage performance. We conducted two workshops with community-based retired women dancers, employing interactive dance and LLM-powered video generation probes in co-design activities. Findings indicate that \deleted{aging-friendly}\added{age-sensitive} adaptations, such as low-barrier keyword input, motion-aligned visual effects, and participatory scaffolds, lowered technical barriers and fostered a sense of authorship. These features enabled retired women to empower their stage, transitioning from passive recipients of stage design to empowered co-creators of performance. We outline design implications for incorporating interactive dance and \added{artificial intelligence-generated content} (AIGC) into the cultural practices of retired women, offering broader strategies for \deleted{aging-friendly}\added{age-sensitive} creative technologies.
\end{abstract}




\begin{teaserfigure}
    \centering
    \includegraphics[width=1.0\textwidth]{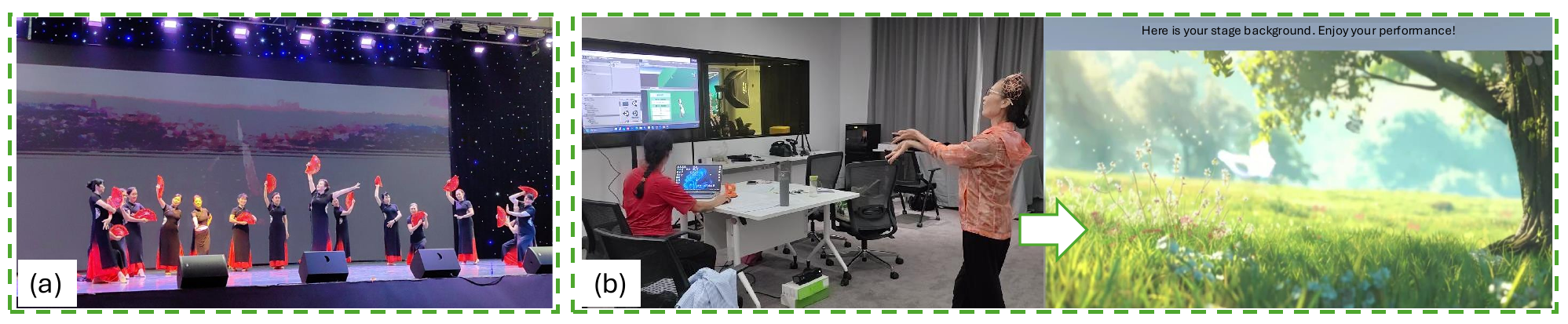}
    \caption{(a) A traditional stage setup for retired women's dance performances, using generic, static backdrops.(b) Our probe in workshop II, StageTailor, enables dynamic co-creation: artificial intelligence (AI) generates semantically aligned backgrounds from keywords, while motion-triggered effects enhance dancer-stage interaction.}
    \Description{(a) A traditional stage setup for retired women's dance performances, using generic, static backdrops.(b) Our probe in workshop II, StageTailor, enables dynamic co-creation: AI generates semantically aligned backgrounds from keywords, while motion-triggered effects enhance dancer-stage interaction.}
    \label{fig:enter-label}
\end{teaserfigure}

\maketitle

\section{Introduction}
In recent years, dance has become a widely adopted leisure activity among retired women in China. With estimates suggesting over 100 million participants, community dance groups and classes at senior universities have become vibrant hubs for this demographic~\cite{Self-cultivation_dance_retiredwomen, senioruniversity, liu2024impact}. In this context, "community" refers to non-professional, predominantly retired women's collectives, bound by shared post-retirement identity, collective social practice, and limited access to professional artistic resources. These activities offer a powerful combination of physical health benefits, enhanced mental well-being, and vital social connection~\cite{physicalbenefits,review_effectsofdance,meaning_of_socialdancing}. As more retired women engage in dance activities and accumulate experience, their aspirations are transcending simple participation and skill acquisition~\cite{liu2024impact}, leading them to seek the expression of artistic sensibilities and beauty through dedicated stage performances~\cite{pursuebeauty}. \added{This transition marks a shift in motivation, from learning-oriented participation to intentional aesthetic creation, which exposes a growing gap between their creative ambitions and the limited stage design resources available in community contexts.}

\deleted{However, despite this enthusiasm, retired women face significant challenges when preparing for and performing on stage.}\added{Translating these creative ambitions into reality, however, remains a fraught process, as retired women encounter a complex interplay of systemic and individual challenges.} These include \textbf{(C1) Lack of professional stage design support}, which restricts their access to professional equipment and technical assistance~\cite{limitedresources}, \textbf{(C2) the age-related decline in physical and cognitive functions}, which can impair movement execution and the recall of complex choreographies~\cite{lackofstandard, physicallimited}, and \textbf{(C3) a persistent digital divide} often hinders their ability to leverage technological tools that could otherwise aid their practice and performance~\cite{Sovhyra2023The}. Collectively, these barriers not only constrain their performance quality but also impede their creative ambitions, highlighting a critical need for accessible, \deleted{aging-friendly}\added{age-sensitive} designs that can empower their artistic journey.

Recent advancements in interactive dance technology offer promising opportunities to enrich stage performances. These systems translate dancers' movements into digital signals through sensors, generating real-time visual feedback that enhances artistic expression~\cite{dannenberg1995model,birringer2004dance,Svenns2020categories}. Prior work shows that these technologies can strengthen stage presence and audience engagement in professional contexts~\cite{Mullis_2013}, suggesting their potential to mitigate challenges related to the lack of professional stage design support (\textbf{C1}) and declining physical precision (\textbf{C2}). However, these systems are largely designed for professional or experimental dance, where precise execution and technical support are assumed~\cite{joshi2021review, Mullis_2013}. Retired women in community dance are typically 50 and above, facing age-related physical, cognitive~\cite{Song2023Physical}, and technological barriers~\cite{Yazdani-Darki2020Older}. 

Moreover, recent progress in Artificial Intelligence Generated Content (AIGC) introduces new possibilities for addressing the digital divide (\textbf{C3}). AIGC systems can create text, images, and video with minimal expertise~\cite{Cao2024A}, lowering reliance on professional tools and budgets. A key enabler here is Large Language Models (LLMs), which allow users to generate content through everyday language or simple keywords~\cite{Wang2024LAVE:}. Prior CHI work has shown that natural-language interfaces and generative artificial intelligence (AI) can help older adults engage in creative, educational, and health-related practices~\cite{Hu2024Engaging, Zhai2024Exploring}. These findings highlight the potential of AIGC and LLM-powered tools to provide retired women with personalized, low-barrier ways of shaping stage visuals.

\added{
At the same time, while AIGC tools are often described as requiring minimal technical expertise, their effective use does not eliminate the need for domain understanding. The quality and relevance of AI-generated outputs remain highly dependent on how prompts are formulated, which presupposes users’ ability to articulate domain-specific intentions.
}

\added{
In the context of community dance, retired women possess embodied knowledge of rhythm, mood, narrative, and aesthetic expression developed through years of practice. However, translating this tacit, experiential knowledge into effective digital input can be challenging. This highlights the need for AI systems that do not replace domain expertise, but instead mediate between embodied artistic understanding and computational generation. This perspective motivates our focus on age-sensitive creative AI mediation: designing AI systems that scaffold the articulation of existing artistic intent, rather than assuming creativity can be automated through generation alone.}

These studies reveal a promising research opportunity, suggesting a powerful, yet unexplored pathway to making complex systems like interactive dance more accessible.

Given this context, our research investigates the following research questions (\textbf{RQs}):

\begin{itemize}
    \item \textbf{RQ1:} How can interactive dance technologies be designed in an \deleted{aging-friendly}\added{age-sensitive} way to support retired women, given their resource constraints and age-related physical characteristics, in enhancing stage performance?
    \item \textbf{RQ2:} How can AIGC and LLM-powered input be integrated into interactive dance systems to lower technical barriers and foster creative participation among retired women?
\end{itemize}

To better understand the unique needs, expectations, and challenges in this context, we first conducted an exploratory workshop with 15 retired female community dancers. Through questionnaires, interviews, and engagement with interactive probes, we identified a foundational tension: while participants faced constraints such as limited resources and age-related physical changes, they held a strong desire for greater creative agency in their stage productions. Specifically, they sought personalized visual backgrounds and more direct input into the stage design to better reflect the artistic intent of their performances. This initial finding highlighted a clear opportunity to explore how technology could support their expressive goals and derived several design considerations (DCs) for a potential interactive system.

Building on these insights, we designed and developed StageTailor, a co-creative system aimed at empowering these dancers. StageTailor integrates LLMs for conversational scene generation, AI-powered video synthesis for creating dynamic backgrounds, and motion capture for real-time interactive effects. We deployed StageTailor in a second workshop with 16 retired women, allowing them to create their own stage visuals. Participants used the system to create backdrops, apply interactive effects, and reflect on their experience through questionnaires, observations, and semi-structured focus group interviews. 

Our findings revealed four core themes that outline both the potential and the challenges for \deleted{aging-friendly}\added{age-sensitive} creative tools: (1) Participants' preferences for keyword-based description were strong; this iterative process lowered cognitive barriers and served as a catalyst for refining ideas, yet participants desired finer control over textual tone and complexity. (2) Their reactions to AI-generated videos were largely positive, though a significant expectation gap emerged in conveying temporal, emotional, and stylistic nuance, highlighting the need for more intuitive communication channels with the AI. (3) Their responses to interactive visual effects were enthusiastic, but also revealed a disconnect between technical interactivity and artistic expression; participants sought effects with greater contextual integration, aesthetic coherence, and customizable variety to match their dance narratives. (4) Their perception of the integrated co-creative experience was profoundly positive; the system facilitated a pivotal shift in their role from passive consumers of pre-made content to active co-creators who could author their stage environment, strengthening their sense of ownership, agency, and emotional investment.

These insights highlight key opportunities for advancing creative AI systems for older adults. By supporting dialogic and multimodal interactions, such as iterative refinement and visual reference, future systems can effectively bridge the gap between embodied expression and digital output. Our work demonstrates how technology can reframe the role of aging populations in creative practices, transforming them from passive recipients into empowered authors of their interactive performances.

In sum, we make the following contributions:
\begin{itemize}
    \item Conceptual Contribution: \deleted{Aging-Friendly}\added{Age-Sensitive} Creative AI Mediation. We articulate and operationalize \deleted{Aging-Friendly}\added{Age-Sensitive} Creative AI Mediation as a novel design framework, demonstrating how AI can effectively redistribute aesthetic control and foster creative agency for retired older adults in the high-stakes domain of stage performance.
    \item Empirical Understanding of Constraints and Empowerment. An empirical understanding of the constraints (resource limitation, low technical self-efficacy) and aspirations of retired women dancers. Our findings surface how participatory scaffolds (low-barrier input, motion-aligned visuals) enable their pivotal transition from passive recipients to empowered co-creators.
    \item Design Insights for Inclusive Creative AIGC Systems. We derive key design implications based on iterative user feedback, particularly highlighting the need for systems to balance creative sophistication with user-centered simplicity and support visual and narrative-oriented control to align AIGC output with nuanced human intent.
\end{itemize}

\begin{table*}[t]

\renewcommand{\arraystretch}{1.5}
\centering
\caption{Taxonomy of interactive dance systems based on users' inputs and systems' outputs, with representative examples illustrating each category.}
\Description{A taxonomy of interactive dance systems, organized by the users' inputs (Movement & Posture, Physiological Features, Props) and the systems' outputs (Geometric, Humanoid, Specific Scenes/Stories). The table includes representative examples for each category.}
\label{table:classification_of_studies}
\small
\begin{tabularx}{\textwidth}{@{} p{3cm} X X X @{}}
\hline
\textbf{\mbox{Inputs (C) $\setminus$ Outputs (R)}} & \textbf{Geometric Shapes} & \textbf{Humanoid Forms} & \textbf{Specific Scenes/Stories} \\ \hline

\textbf{Movement} & 
\textbf{9 studies} \cite{kepner1997, Qian2004, barry2005, james2006, jung2011, Mullis_2013, seo2017art, wu2019, masu2022} \par
\textit{\footnotesize Example: Motion capture projects dynamic visuals responding to dancers' movement, from lines to organic ink cloud forms~\cite{Mullis_2013}.} & 
\textbf{3 studies} \cite{seo2017art, fdili2019, masu2022} \par
\textit{\footnotesize Example: Full-body tracking creates visual effects that mirror and amplify the dancers' movements~\cite{masu2022}.} & 
\textbf{5 studies} \cite{kepner1997, Qian2004, barry2005, Park2006, wu2019} \par
\textit{\footnotesize Example: Dancer's movements control an emperor's expression, with thunder, lightning, and light effects creating narrative responses~\cite{kepner1997}.} \\ \hline

\textbf{Physiology} & & & 
\textbf{2 studies} \cite{Park2006, fdili2019} \par
\textit{\footnotesize Example: Biometric sensors transform physiological data into immersive audiovisual narratives that enhance physical storytelling~\cite{fdili2019}.} \\ \hline

\textbf{Props} & 
\textbf{2 studies} \cite{Latulipe2008, seo2017art} \par
\textit{\footnotesize Example: Sensor-equipped pillows transform prop interactions into projected grid patterns that illustrate the narrative~\cite{seo2017art}.} & & 
\textbf{1 study} \cite{Latulipe2008} \par
\textit{\footnotesize Example: Wireless mice allow dancers to control dynamic projections, creating "flying origami" and other geometric narratives~\cite{Latulipe2008}.} \\ \hline
\end{tabularx}
\end{table*}

\section{RELATED WORK}

\subsection{Challenges Faced by Retired Women in Dance Stage Performance}
\label{subsec:Challenges Faced by Retired Women in Dance Stage Performance}
In China, dance has become a widespread post-retirement activity among women, serving as a means to enrich daily life and pursue aesthetic expression through stage performance. Studies estimate that over 100 million retired women in China have joined or organized public dance groups after leaving the work force~\cite{Self-cultivation_dance_retiredwomen}. These participants often take part in dance courses offered by local communities or senior universities~\cite{senioruniversity, liu2024impact}. Such involvement promotes physical fitness, supports mental well-being, and strengthens social connectedness~\cite{Self-cultivation_dance_retiredwomen, physicalbenefits,review_effectsofdance,meaning_of_socialdancing}. As their experience deepens, many retired women move beyond leisure participation to view stage performance as a chance to showcase learning and self-expression~\cite{liu2024impact}. They increasingly aspire to create aesthetically pleasing performances, pursuing beauty and expressive impact~\cite{pursuebeauty}. What begins as class task thus often evolves into a more intentional effort to refine the artistic quality of their performances. \added{This shift reflects a growing creative ambition among retired women dancers: as they accumulate experience and confidence, stage performance is no longer perceived merely as an outcome of learning, but as an expressive medium through which aesthetic intent, emotional atmosphere, and group identity are communicated. 
However, this evolution in motivation is not matched by corresponding access to professional stage design resources or creative tools, creating a tension between rising artistic aspirations and limited means of realization.

Building on this tension, prior studies and observations identify several persistent challenges (Cs) that constrain the quality and expressiveness of retired women’s stage performances:}

\deleted{However, despite their enthusiasm, retired women often face several persistent challenges that limit the quality and expressiveness of their stage performances. Prior studies and observations point to several persistent challenges (Cs):}

\begin{itemize}
    
    \item C1: Lack of professional stage design support. Performances typically occur on community stages lacking professional designers, equipment, or technical staff. Unlike formal dance troupes that work with lighting designers, visual artists, and stage technicians, retired women must rely on improvised costumes and minimal backstage support. Without dedicated assistance for planning or operating stage effects, their performances are constrained in visual presentation, limiting the expressiveness and impact~\cite{limitedresources}.
    
    \item C2: Age-related decline. Natural age-related changes in physical and cognitive functions affect performance quality. Declines in strength and flexibility can reduce the precision of movements, while memory decline makes recalling long sequences difficult, impacting group synchronization \cite{limitedresources,physicallimited,lackofstandard}. These challenges stem from the natural effects of aging, not a lack of motivation, but they constrain the dancers' ability to achieve the polished performances they aspire to. 
    
    \item C3: The technology gap. While digital tools for stage performance are increasingly available, older dancers often struggle to access them. Barriers range from physical difficulties with small interfaces to a lack of confidence or social support in using new tools \cite{Sovhyra2023The, Martins2025From,Wildenbos2019Mobile,Wilson2021Barriers,Vaportzis2017Older,Tian2024Benefits,Pywell2020Barriers}. This creates a significant gap between the potential of technology and its actual usability for this group, preventing them from leveraging innovations that could otherwise enhance their performances.
\end{itemize}

Collectively, these challenges faced by retired women dancers are not due to a lack of creativity, but to a combination of scarce resources, age-related changes, and a technology gap. This necessitates \deleted{aging-driendly}\added{age-sensitive} designs that are both technically accessible and artistically empowering.

\subsection{The Potential of Interactive Dance in Addressing C1 and C2}

Building on the challenges faced by retired women dancers, we asked how technology might be adapted to empower them in their stage performances. Inspired by recent work showing how interactive technologies can enrich professional dance performances~\cite{Mullis_2013, joshi2021review}, we propose that interactive dance, where a dancer's movements shape the surrounding visual or auditory environment in real-time~\cite{Politis1990review, Mullis_2013}, offers a compelling and feasible solution.

While the field has explored various interaction modalities like music adaptation and lighting control~\cite{Svenns2020categories}, these often require technical infrastructure that is impractical for retired women's community performances. These groups often lack professional stage design support (C1), performing on local stages where background screens are often the only available equipment and music is fixed in advance \cite{limitedresources}. Within these constraints, interactive visual feedback emerges as the most feasible and impactful modality, as it can be projected onto existing screens without disrupting established choreography or rehearsal routines.
In this context, interactive visuals can directly address key challenges faced by retired women dancers:

\begin{itemize}
    \item Compensating for the lack of professional stage design support (C1): On community stages lacking professional designers or technical staff, interactive systems can interpret movements into responsive backgrounds and effects. This augments the performance and enhances the overall presentation, effectively compensating for the absence of dedicated stage staff.

    \item Mitigating the Impact of Age-Related Decline (C2): As physical strength and coordination naturally decrease with age, maintaining precision can become a source of stress. By creating a captivating visual layer that responds to the dancers' collective movement, interactive effects can shift the audience’s focus from individual accuracy to the overall expressive quality of the performance, fostering a more holistic and engaging experience.
\end{itemize}

However, the technology gap (C3) remains unresolved. Existing interactive dance systems are primarily designed for professional contexts, assuming a level of technical expertise and support that is misaligned with the realities of community dance \cite{Mullis_2013, joshi2021review}. This gap highlights the need to rethink how interactive visuals can be adapted for aging dancers, empowering them to harness the expressive potential of these technologies.

To inform our approach, we conducted a literature review of 12 representative studies to create a taxonomy of interactive dance types. This review served a dual purpose: (1) to structure our introduction of key concepts to participants and (2) to guide the design of our interactive system. We classified the studies based on two dimensions: interaction mode (input) and visual feedback (output). \autoref{table:classification_of_studies} presents this classification with representative examples.

\begin{table*}[t]
\centering
\small
\caption{Participants of the Workshop I organized by groups (G1–G4). The table lists individual ages, dance experience levels, community performance frequencies, and team types.}
\Description{This table details the demographic and performance backgrounds of participants in Workshop I, comprising 15 retired women aged 51–67 (mean: 56.27). All participants had at least two years of dance experience, with performance frequencies ranging from monthly to biannual. The table also distinguishes team types, including senior university groups and community-based ensembles, highlighting variations in performance opportunities. Notably, some participants engaged in both educational and informal dance settings, reflecting the diverse organizational structures shaping their artistic practices.}

\label{tab:Workshop I participants}
\begin{tabularx}{\textwidth}{@{} p{1.5cm} p{1.5cm} p{1.5cm} p{3cm} p{3.5cm} X @{}}
\toprule
\textbf{Group} & \textbf{ID} & \textbf{Age} & \textbf{Years of Dancing} & \textbf{Freq. of Performances} & \textbf{Team Type} \\
\midrule
\multirow{5}{*}{G1} 
 & P1 & 57 & Over 3 years & once every 6 months & Senior university / Community \\
 & P2 & 52 & Over 3 years & Every 2--3 months & Senior university \\
 & P3 & 57 & Over 3 years & Every 2--3 months & Senior university / Yue Women's Academy \\
 & P4 & 60 & Over 3 years & Every 2--3 months & Senior university / Community \\
\midrule
\multirow{4}{*}{G2} 
 & P5 & 67 & Over 3 years & Every 2--3 months & Community \\
 & P6 & 60 & Over 3 years & Every 2--3 months & Community \\
 & P7 & 58 & Over 3 years & Every 2--3 months & Community \\
 & P8 & 52 & Over 3 years & Every month & Community \\
\midrule
\multirow{4}{*}{G3} 
 & P9 & 56 & Over 3 years & Every month & Senior university / Community \\
 & P10 & 52 & 2--3 years & Rarely (end-of-semester recital) & Senior university \\
 & P11 & 51 & Over 3 years & Rarely (end-of-semester recital) & School \\
\midrule
\multirow{2}{*}{G4} 
 & P12 & 55 & Over 3 years & once every 6 months & Senior university / Community \\
 & P13 & 55 & Over 3 years & once every 6 months & Community \\
 & P14 & 57 & 2--3 years & Every 2--3 months & Community \\
 & P15 & 56 & Over 3 years & Every 2–3 months & Senior university / Community \\
\bottomrule
\end{tabularx}
\end{table*}

\subsection{The Potential of LLM-powered AI Video Generation in Addressing C3}

Recent advances in AIGC make it possible to generate text, images, video, and audio with minimal effort~\cite{Cao2024A, Hua2024Generative}, impacting professional creative domains like stage design and visual arts \cite{Cao2024A, Hua2024Generative, Liu2024AIGC, AIGCSilkroad}. While many of these applications assume professional contexts, this technology holds immense potential to bridge the technology gap for non-professional users:

\begin{itemize}
    \item Bridging the technology gap (C3). Most applications assume professional contexts with budgets, equipment, and technical expertise—conditions absent from community performances, which often rely on nothing more than a single background screen. This gap suggests the potential of AIGC to democratize stage creation for non-professional users. A key enabler here is LLMs, which allow users to describe ideas in everyday language and automatically turn them into generative outputs~\cite{Wang2024LAVE:}. Tools such as LAVE demonstrate how plain-language input can replace technical editing commands.
\end{itemize}

Recent work shows that LLMs can also empower older adults. For instance, the CommUnity AI toolkit enabled seniors to co-create community renewal projects through simple descriptions ~\cite{Hu2024Engaging}. Similar studies in CHI confirm that natural-language interfaces help older adults participate in creative, educational, and health activities~\cite{Khamaj2025AI-enhanced, Zhai2024Exploring, liuchao}. This body of work points to the strong potential of \deleted{aging-driendly}\added{age-sensitive} AIGC tools to provide older adults with simple, personalized ways to engage in co-creation.

\section{Workshop I: Understanding Retired Women}
\label{sec:Workshop I}
\begin{figure*}[t]
    \centering
    \includegraphics[width=1.0\textwidth]{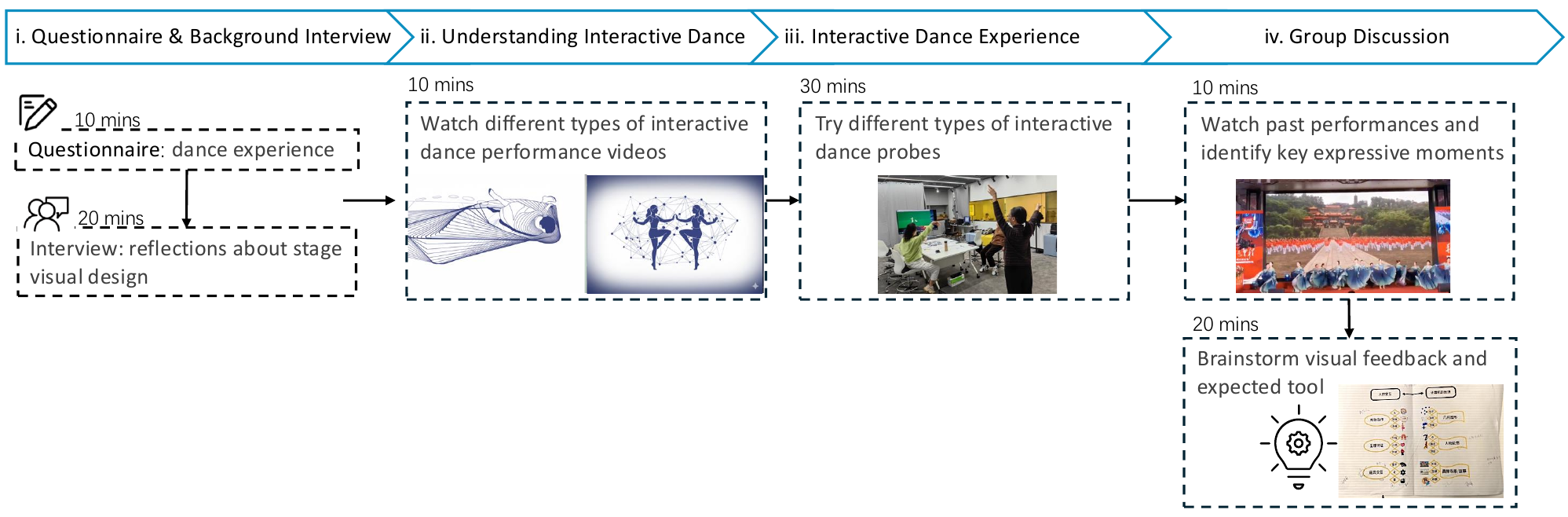}
    \caption{Procedure of the Workshop I with 15 participants divided into four groups (3-4 participants per group) to understand retired women’s perspectives on current stage performances and interactive dance. The study followed a four-phase structure: (i) a preliminary questionnaire and background interview; (ii) an introduction to interactive dance concepts; (iii) a hands-on experience with interactive dance probes; (iv) a concluding group discussion to synthesize their experiences and ideas.}
    \Description{Procedure of the Workshop I with 15 participants divided into four groups (3-4 participants per group) to understand retired women’s perspectives on current stage performances and interactive dance. The study followed a four-phase structure: (i) a preliminary questionnaire and background interview; (ii) an introduction to interactive dance concepts; (iii) a hands-on experience with interactive dance probes; (iv) a concluding group discussion to synthesize their experiences and ideas.}
    \label{fig:US1 process}
\end{figure*}

To understand retired women's perspectives on their current stage performances and their attitudes towards the integration of interactive visual feedback, we designed a set of activities in Workshop I. As illustrated in~\autoref{fig:US1 process}, there are four main phases: (i) a preliminary questionnaire and background interview; (ii) an introduction to interactive dance concepts; (iii) a hands-on experience with interactive dance probes; (iv) a concluding group discussion to synthesize their experiences and ideas.

\subsection{Participants}
 We recruited 15 retired women (ages 51–67, M = 56.27, SD = 4.31) through snowball sampling from three retirement communities. Inclusion criteria required participants to have at least two years of dance experience and a minimum of two public performances. To ground the activities in their shared experience, we organized participants into four groups (three groups of four, one of three), ensuring that members of each group had previously performed choreography together. 
 Table~\ref{tab:Workshop I participants} provides the detailed participant information.

\subsection{Apparatus and Materials}
To facilitate participants' understanding and engagement with interactive dance, we prepared several materials and technical components used during the study process. These resources were integrated into different phases of the study as outlined below.

\subsubsection{Inspirational materials}
To introduce participants to the \textit{six core categories of interactive dance} from our taxonomy (\autoref{table:classification_of_studies}), we curated a diverse set of stimuli. This set included \textit{inspirational videos of professional performances}, \textit{images from prior literature}, and \textit{a series of hands-on probes} developed with Unity~\cite{unity2019}, the RAM Dance Toolkit~\cite{phdthesis}, and Kinect~\cite{kinect}. The videos and images provided concrete examples for each category, while the probes offered a simplified, tangible way for participants to experience the underlying dynamics firsthand. This approach was designed to ground abstract concepts in both inspirational examples and direct, bodily experience. 

\begin{itemize}
    \item \textbf{Movement \& Posture → Geometric Shapes.} For this category, we provided two videos and one probe. The videos demonstrated how geometric lines and shapes can be projected directly onto dancers and the stage floor, dynamically responding to their movements~\cite{Mullis_2013, examplevideomullis}, or how a dancer’s gesture could trigger bursts of fireworks~\cite{examplevideo1}. The probe was a real-time system that captured participants’ body movements via Kinect and rendered them into evolving geometric particles and lines.
    
    \item \textbf{Movement \& Posture → Humanoid Forms.} For this category, we provided one video and one probe. The video showed multiple human silhouettes that were projected to collectively form shapes such as characters or animals, and these silhouettes then responded to and interacted with elements in the background scene~\cite{examplevideorumenglai}. The probe mapped participants’ shadows to reveal hidden photographic layers behind a white screen, encouraging them to use their bodies as “brushes” to uncover imagery.
    
    \item \textbf{Movement \& Posture → Specific Scenes/Stories.} For this category, we provided one video and one probe. The video showed dancers inserted into cityscapes and traditional ink paintings, where their gestures manipulated cars or weather~\cite{examplevideorumenglai}. The probe allowed participants’ motions to drive a 3D avatar accompanied by a dog, whose behaviors (e.g., running, following) responded to the dancer’s movements within a virtual environment.
    
    \item \textbf{Physiological Features → Specific Scenes/Stories.} For this category, we provided two examples through images. The first one demonstrated how biometric sensors (e.g., Mio armband, heart-rate monitors) could control scene properties such as zoom and blur, transforming performers’ physiological states into aesthetic variations. This example helped participants imagine interaction beyond visible movement~\cite{fdili2019}. The second image, which we explained to participants, depicted a dancer's movements influencing the expressions of an emperor and the atmosphere, serving as an example of how gestures could drive a narrative~\cite{Park2006}.
    
    \item \textbf{Props → Geometric Shapes.} For this category, we provided one image and one probe. The image demonstrated how dancers used tangible objects like pillows to modulate grid patterns, extending their bodily expression~\cite{seo2017art}. The probe adapted this concept by letting participants dance with a handheld mouse, which generated moving geometric trails in response to prop gestures.
    
    \item \textbf{Props → Specific Scenes/Stories.} For this category, we provided one probe. The probe allowed participants to use a handheld mouse to interact with a naturalistic scene: moving across a virtual meadow while scrolling the mouse wheel launched paper planes, evoking playful narrative responses to prop manipulation.

\end{itemize}

\subsubsection{Participants' Performance Videos.} Prior to the workshop, each group submitted a video of a past performance.  These videos served as personal and familiar artifacts for grounding the co-design activities, enabling participants to reflect on how interactive technologies could be applied to their own choreography.

\subsection{Procedure}
As shown in~\autoref{fig:US1 process}, the 100-minute workshop was conducted in a quiet, interruption-free room. \added{While live performances often take place in dynamic, noisy environments, this controlled, interruption-free setting was intentionally chosen for Workshop I. It aimed to create a "safe space" that fostered deep reflection and ensured psychological comfort, allowing participants to share personal stories and creative frustrations without external distraction.} Each group sat together around a table to facilitate discussion and collaboration.

\textbf{Part 1: Background and Context.} Participants completed a questionnaire covering their demographic information, dance experience, and views on stage performance. This was followed by semi-structured interviews focusing on their opinions regarding current stage visuals and backgrounds.

\textbf{Part 2: Introduction to Interactive Dance.} The researchers presented a curated set of interactive dance videos and explained the underlying technologies, their applications, and relevance to theatrical settings. This provided participants with foundational knowledge and conceptual grounding.

\textbf{Part 3: Hands-on Exploration with Probes.} Participants interacted with a series of interactive dance prototypes in a flexible and exploratory setting. Researchers offered guidance as needed, while encouraging open-ended engagement. These hands-on experiences exposed participants to different interaction styles and feedback mechanisms, generating valuable insights for later analysis.

\textbf{Part 4: Co-design and Reflection.} Participants engaged in a brainstorming session. They watched videos of their own performances, identified key moments, and proposed enhancements using visual or interactive feedback.   Their designs and ideas were recorded and refined through group discussion, allowing researchers to better understand how participants envisioned the integration of interactive dance into their practices.

\subsection{Data Analysis}

\deleted{Data sources included questionnaire responses, interview transcripts, observational notes, photographs, and participant-created design sketches. All audio and video recordings were verbatim transcribed. We conducted thematic analysis, employing open coding with a combined deductive and inductive approach to develop the initial coding framework. To ensure coding consistency, two researchers independently coded segments of the data. They met iteratively to discuss all preliminary codes and resolve discrepancies, reaching consensus on a finalized coding manual. This consensus-driven manual was then strictly applied to the remaining data, ensuring the reliability of the resulting themes. This process established two primary themes: (1) dissatisfaction with and limitations of current stage backgrounds, and (2) expectations for interactive visual feedback. Sub-themes were constructed through keyword analysis and grouped into a hierarchical framework. Coding discrepancies were resolved through iterative discussion to ensure consistency and reliability.}

\added{Data sources included questionnaire responses, interview transcripts, observational notes, and participant-created sketches. All recordings were transcribed verbatim. We analyzed the data using reflexive thematic analysis~\cite{forbes2022thematic, braun2022starting}, following an inductive coding process. To ensure depth and rigor, two researchers first coded the data independently to capture diverse perspectives. They then met regularly to discuss their initial codes, sharing personal reflections and comparing interpretations. Rather than simply resolving "disagreements" to reach a fixed manual, these discussions served to enrich our understanding and helped us generate more nuanced themes. This collaborative and reflexive process led to two primary themes: (1) dissatisfaction with and limitations of current stage backgrounds, and (2) expectations for interactive visual feedback. These themes were further refined into a hierarchical framework that reflects the participants' creative aspirations.}

\subsection{Findings}
We present our key findings on two themes. The first theme details participants' dissatisfaction with current stage backdrops, highlighting a significant gap between their artistic aspirations and the practical constraints they face. The second theme outlines their aspirations for future systems, focusing on how they envision interactive technology empowering them as creative partners in the design process.

\begin{figure*}[t]
    \centering
    \begin{subfigure}[t]{0.4\textwidth}
        \includegraphics[width=\linewidth]{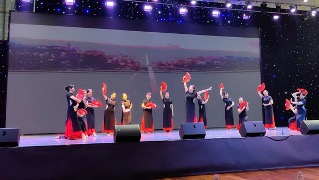}
        \caption{Lacking thematic relevance: A traditional Chinese fan dance in front of an irrelevant cityscape backdrop.}
        \label{fig:home-delivery}
    \end{subfigure}%
    \hspace{1cm}
    \begin{subfigure}[t]{0.4\textwidth}
        \includegraphics[width=\linewidth]{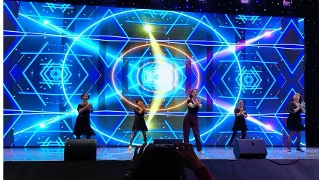}
        \caption{Distracting visual overload: Overuse of neon grids and particles overwhelms a Latin dance performance.}
        \label{fig:orders}
    \end{subfigure}%
        \\
    \begin{subfigure}[t]{0.4\textwidth}
        \includegraphics[width=\linewidth]{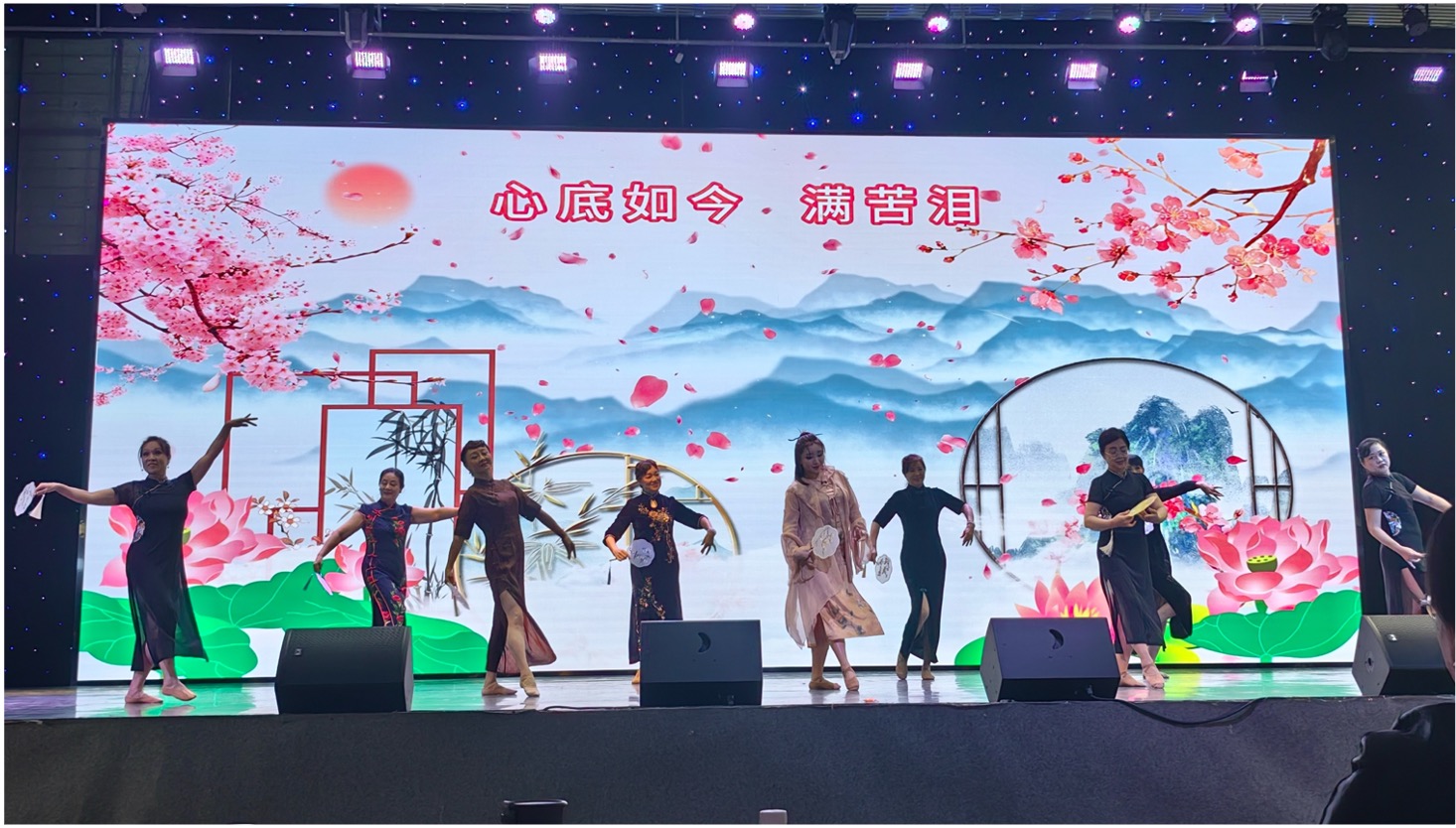}
        \caption{Poor color coordination: Background and lighting blend with costumes, obscuring dancers' faces and movements.}
        \label{fig:orders-details}
    \end{subfigure}%
    \hspace{1cm}
    \begin{subfigure}[t]{0.4\textwidth}
        \includegraphics[width=\linewidth]{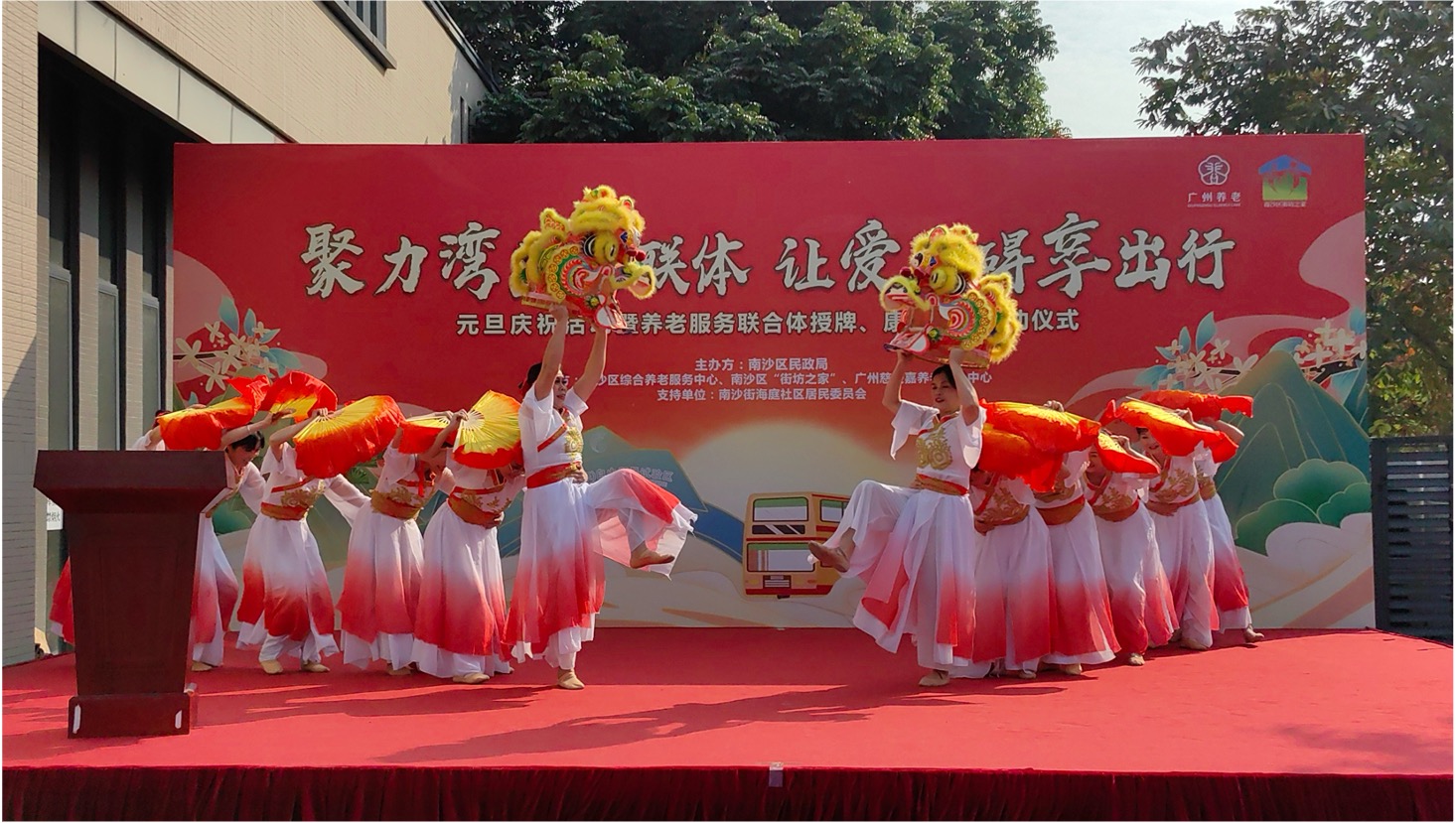}
        \caption{Limited visual support: Outdoor performance relies on a static backdrop without dynamic effects.}
        \label{fig:account}
    \end{subfigure}
    \caption{Example scenes from real-world performances illustrating common issues with current stage backdrops in retired women's dance. These backdrops are usually selected by group leaders from online videos or static images that loosely match the dance theme or regional style (e.g., grasslands for Xinjiang dances, classical paintings for fan dances). In some cases, participants attempt to manually stitch together regional photos for promotional purposes, though with limited tools and skills.}
    \Description{Example scenes from real-world performances illustrating common issues with current stage backdrops in retired women's dance. These backdrops are usually selected by group leaders from online videos or static images that loosely match the dance theme or regional style (e.g., grasslands for Xinjiang dances, classical paintings for fan dances). In some cases, participants attempt to manually stitch together regional photos for promotional purposes, though with limited tools and skills.}
    \label{fig:Current Stage}
\end{figure*}

\subsubsection{Dissatisfaction and Limitations of Current Stage Backdrop} Participants expressed widespread dissatisfaction with their existing stage visuals, citing a disconnect from their performances and a lack of agency in the creative process (\autoref{fig:Current Stage}).

\textbf{Thematic and embodied disconnect between dance and backdrop.} A primary source of frustration, cited by 13 participants, was the minimal connection between the background visuals and the dance itself. Visuals were often only loosely related to the performance's theme and failed to reflect the dancers' movements, thereby failing to enhance the overall stage effect. One participant (P3) specifically noted that in some cases, overly complex or visually cluttered backgrounds made it difficult for the audience to clearly see the dancer's movements from the front of the stage, causing distraction and diminishing the visual focus on the performer. Participants in G2 similarly observed that when their dance included powerful or expressive movements, the background visuals failed to mirror or amplify that energy. This resulted in a perceived disconnect that undermined the performance's artistic impact.

\textbf{Lack of time and expertise to create appropriate stage backgrounds.} While participants were dissatisfied with the current stage backgrounds, they (N=8) felt powerless to improve their stage visuals due to a lack of technical expertise, time, and resources. Backgrounds were often hastily selected by organizers or instructors shortly before performances, typically by browsing online for superficially matching content. For instance, a Xinjiang-style dance might be paired with a generic grassland video that lacks cultural specificity or alignment with choreography. While some (G1) attempted to create their own visuals by editing images, they found the process \textit{''time-consuming''} and “technically demanding.” This sentiment was succinctly captured by P2: \textit{''This is the best we can do with what we have.''} Consequently, most eventually gave up on contributing to stage design, focusing instead on dance rehearsal. Stage visuals remained outside their creative reach.

\textbf{Performance fatigue from static and repetitive visuals.} The practical constraints often led to the repeated use of the same dance and backdrop for extended periods, sometimes for one to two years. Five participants described a growing sense of fatigue with this repetition. \textit{''The first few times, the backdrop seemed fine,''} one participant (P7) noted, \textit{''but after a year, we and the audience were tired of it.''} This feeling of stagnation deepened their dissatisfaction, not just with the visuals but with the overall lack of freshness in their stage presentation. The inability to easily update or evolve the stage effects left them feeling stuck in a creative rut, reinforcing their disengagement from the design process.

\subsubsection{Expectations for Co-Creative and Responsive Visuals}
Through the study, participants expressed their expectations regarding the use of interactive dance technologies with visual feedback to enhance the stage backdrop. Their aspirations centered on two key areas: their desired role in the pre-performance design process and their specific expectations for the visual feedback itself.

\textbf{Desire for Pre-Performance Co-Creation.} Contrary to a vision of real-time improvisation on stage, nearly all participants (N=14) expressed a strong preference for engaging in the creative design process before the performance. This preference was shaped by three factors:
\begin{itemize}
    \item \textbf{Practical Constraints of Live Interaction.} Participants (N=14) expressed a clear preference for designing visual elements prior to the performance, rather than interacting with computers in real-time on stage. Participants from G3 noted that, unlike professional dancers in interactive performances, they were unable to engage in spontaneous digital interaction during live shows. This limitation was not due to a lack of interest in improvisation, but rather the high cognitive demands of focusing on choreographed movements and spatial coordination during group dances. One (P9) emphasized that spontaneous actions might disrupt the overall group synchronization, making pre-designed visuals a more practical choice.
    \item \textbf{Interest in Participatory Design.} Participants (N=5) expressed a strong willingness to be involved in designing the interaction methods used to generate visual backdrops. Rather than relying on fixed templates or designer-driven choices, they preferred to select interaction styles aligned with the unique theme, rhythm, and mood of each dance. During the study, common modes of interaction included movement-triggered visuals and the use of physical props, demonstrating their openness to different creative approaches.
    \item \textbf{Desire for Autonomy and Personalization.} A recurring theme was the desire for more creative control over visual content. Participants (N=7) hoped to contribute their own materials --- from specific imagery to the timing of visual transitions --- to reflect their unique interpretation of the dance. Participants from G1 also articulated specific ideas about which segments of the performance should be emphasized visually. This reflects a broader wish for co-ownership in the creative process, allowing for more expressive and personalized stage designs.
\end{itemize}

\textbf{Expectations for Expressive Visual Feedback.} Participants (N=13) had specific expectations for the quality and behavior of visual feedback, emphasizing the need for visuals that were both meaningful and supportive of their performance.
\begin{itemize}
    \item \textbf{Contextual and Thematic Alignment.} There was a strong consensus (N=13) that visuals must be thematically and narratively consistent with the dance. Story-driven backgrounds were especially favored, as they facilitated narrative comprehension and immersive atmosphere. This reflects a desire for semantic coherence between the visual and performative elements, allowing audiences to better grasp the intended emotion and message.

    \item \textbf{Embodied and Motion-Responsive Effects.} Many participants (N=9) appreciated visual effects that responded to specific dance movements, describing these as making the performance \textit{''come alive.''} They found such interactions engaging and expressive, especially when effects followed the rhythm or gestures of the dancers. This preference indicates a desire for temporal synchronization between movement and visuals, enhancing the sense of connection between performer and digital backdrop.
    
\end{itemize}

These findings suggest a clear design trajectory: retired women dancers seek creative tools that empower them as authors in the pre-production phase, enabling them to design stage visuals that are thematically coherent, responsively embodied, and visually clear.

\subsection{Design Considerations}
Synthesizing the findings from Workshop I, we derived three design considerations (DCs) to guide the development of our co-creative system for Workshop II. These DCs address participants' key frustrations with existing tools and their aspirations for a more empowering creative process.

\begin{itemize}
    \item \textbf{DC1: Support Low-Barrier Entry and Broad Applicability.} The tool should be applicable to a wide range of dance styles and intuitive enough for users with diverse technical backgrounds. Unlike interactive dance systems in existing studies~\cite{Mullis_2013,seo2017art,wu2019,fdili2019,masu2022}, which are typically custom-built by expert teams for specific performances, our tool is intended for broader application and cannot provide personalized services in advance (C1). Furthermore, participants in our focus group emphasized the importance of ensuring visual feedback is appropriately matched to both dance and setting. Therefore, the tool must be adaptable across various dance contexts and straightforward to use.

    \item \textbf{DC2: Foster a Sense of Connection Through Embodied Interaction.} A central theme in our findings was the perceived disconnect between static backdrops and the dynamic, embodied nature of dance. To address this, the system should enable a meaningful link between movement and visual feedback. This involves providing effects that can respond to the tempo, gestures, or spatial positioning of the dancers, creating a dynamic interplay that reinforces the performance's narrative and emotional tone, making the visuals an integral part of the choreography rather than a decorative layer.

    \item \textbf{DC3: Empower Creative Agency Through Customization and Ownership.} Participants consistently expressed a desire to move from being passive performers to active authors of their stage aesthetic. They valued personalization and wished to infuse their unique interpretations into the visual design. Therefore, the tool should provide a flexible creative space where users can not only select and modify visual effects but also feel a sense of ownership over the final output. This empowers their creative agency, allowing the stage design to be an authentic reflection of their artistic vision.
\end{itemize}

\begin{figure}[t]
    \centering
    \includegraphics[width=1.0\linewidth]
    {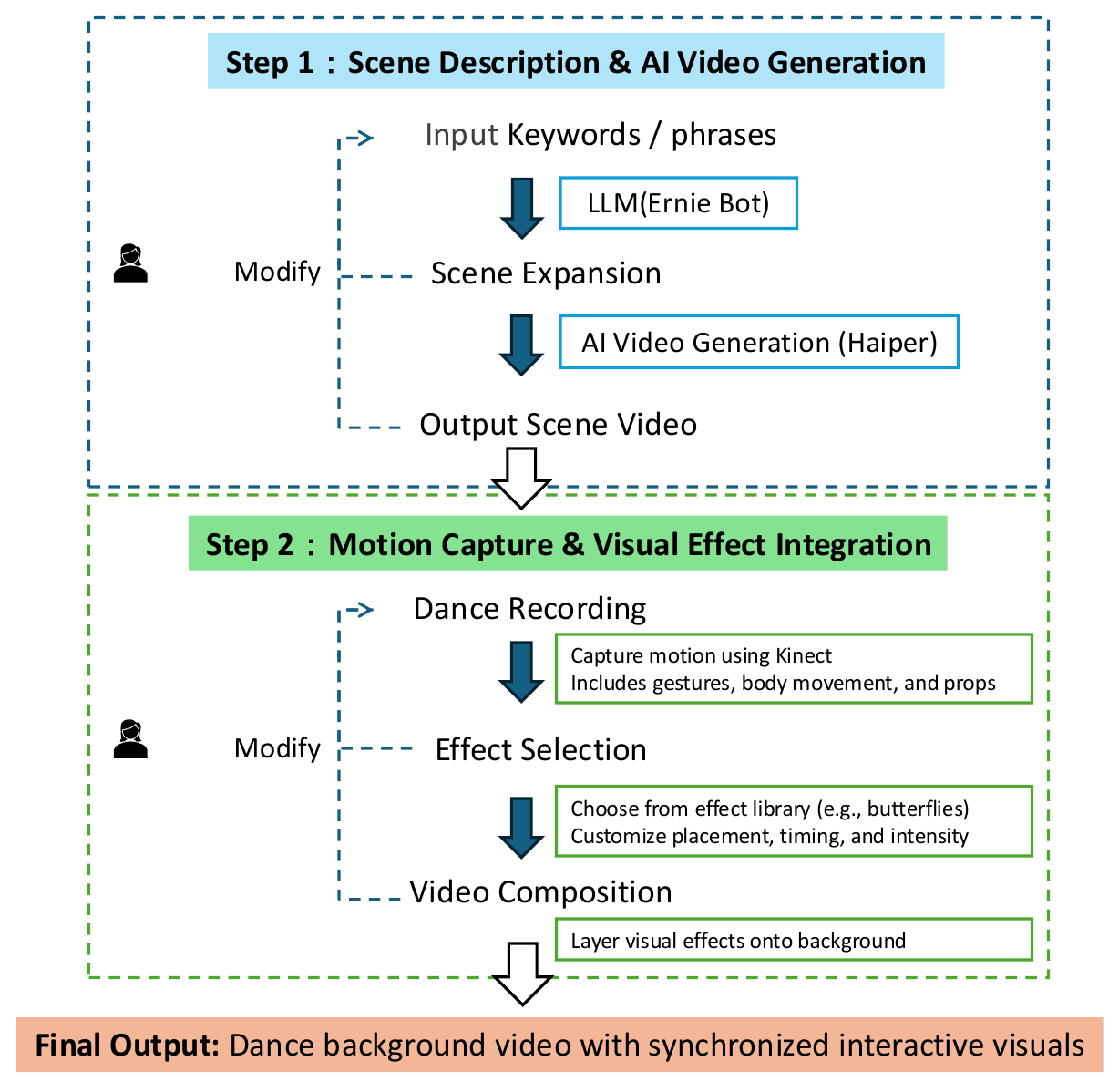}
    \caption{Workflow of StageTailor, a two-step background creation tool designed for retired dancers. In Step 1, users describe a desired scene using keywords, which are expanded by LLM and used to generate a background video via AI text-to-video synthesis. In Step 2, users record their dance movements and select motion-responsive visual effects from a system library. These effects are synchronized with the movements and integrated into the generated video.}
    \Description{Workflow of StageTailor, a two-step background creation tool designed for retired dancers. In Step 1, users describe a desired scene using keywords, which are expanded by LLM and used to generate a background video via AI text-to-video synthesis. In Step 2, users record their dance movements and select motion-responsive visual effects from a system library. These effects are synchronized with the movements and integrated into the generated video.}
    \label{fig: Workflow of StageTailor}
\end{figure}

\section{Probe Design: StageTailor}
\label{sec:stagetailor}
Based on the DCs form workshop I, we designed StageTailor as a research probe to examine how \deleted{aging-driendly}\added{age-sensitive} adaptations of
LLM-powered video generation and interactive visuals could scaffold creative
agency in this context. We translated the DCs into concrete interaction mechanisms and the probe was intentionally designed with a two-step workflow, directly addressing their expressed needs for simple input (\textbf{DC1}), creative authorship (\textbf{DC3}), and a meaningful connection between movement and visuals (\textbf{DC2}).

\begin{figure*}[t]
    \centering
    \includegraphics[width=1.0\textwidth]{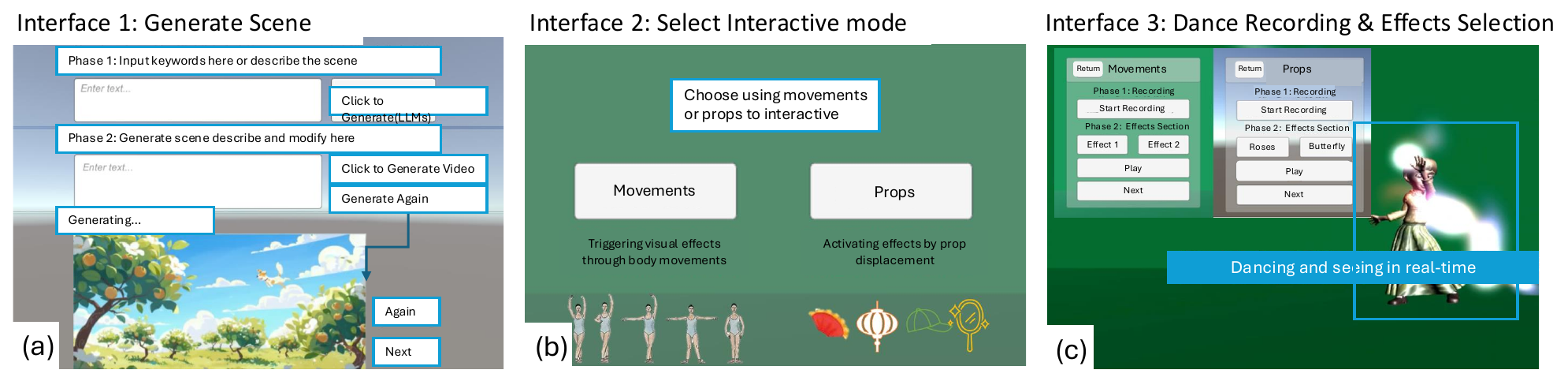}
    \caption{Several example interfaces of Stagetailor: (a) AI Scene Generation Interface, allowing users to input keywords for automated video scene creation, with options for direct generation or stepwise refinement. (b) Interaction Mode Selection, offering body-motion and prop-based control schemes, including a 4s calibration countdown. (c) Dance Effects Recording Panel, featuring real-time motion capture, visual effect presets, and terminal feedback for seamless creative iteration.}
    \Description{Several example interface of Stagetailor: (a) AI Scene Generation Interface, allowing users to input keywords for automated video scene creation, with options for direct generation or stepwise refinement. (b) Interaction Mode Selection, offering body-motion and prop-based control schemes, including a 4s calibration countdown. (c) Dance Effects Recording Panel, featuring real-time motion capture, visual effect presets, and terminal feedback for seamless creative iteration.}
    \label{fig:User interface prototype}
\end{figure*}

\subsection{Design Goals}
StageTailor was shaped around three goals derived directly from the DCs:

\begin{itemize}
  \item \textbf{G1: Lower the barrier to visual creation (DC1).}
  The system should support simple input—such as keywords or short descriptions—
  and help participants articulate their intentions without requiring technical
  vocabulary.

  \item \textbf{G2: Support creative authorship and revision (DC3).}
  Participants should be able to review, refine, and personalize generated
  outputs, ensuring they retain ownership over the creative process.

  \item \textbf{G3: Enable meaningful movement–visual connections (DC2).}
  Interactive effects should respond expressively to body movement, enhancing
  the performance and supporting embodied creative expression.
\end{itemize}

\subsection{Two-Step Workflow}
In response to these goals, we designed a two-step workflow that mirrors how
participants naturally conceptualized their stage needs: first imagining the
scene, then imagining how movement might activate it. \autoref{fig: Workflow of StageTailor} illustrates the workflow.

\subsubsection{Step 1: Describe a Scene and Generate Video (G1, G2).}
This initial step focuses on amplifying creative intent by lowering the barrier to content creation. It enables users to generate customized background videos from simple text inputs, transforming their abstract ideas into unique, tangible visual assets. This capability is crucial, as traditional searching for stock footage would force aesthetic compromise and fail to meet the participants' highly specific, situated stage needs. This necessity grounded our choice of integrating two AI techniques, LLMs~\cite{radford2019llm} and text-to-video generation~\cite{Choi2016, liu2020, liu2021}, which together can expand incomplete descriptions and generate tailored visuals without requiring professional skills.

\begin{enumerate}
    \item \textbf{From Keywords to Scene Description:} Users can either input keywords directly or interact with the LLM in natural language. An integrated LLM then expands these inputs into a rich, scene description, which users can iteratively review and refine. This scaffolds the creative process, helping them articulate their artistic intent without needing technical vocabulary.

    \item \textbf{From Text to Video:} Once finalized, a text-to-video model synthesizes it into a unique background video. This generation process is paramount, as it ensures that the visual output is thematically and emotionally perfectly aligned with their vision, thus maximizing their creative authorship over the final asset—a goal unattainable through the retrieval of generic stock content.
    
\end{enumerate}

We used \textit{Ernie Bot}\cite{erniebot}, a Chinese-language LLM, and \textit{Haiper}\cite{haiper} for text-to-video generation.

\begin{figure*}[t]
    \centering
    \includegraphics[width=1.0\textwidth]{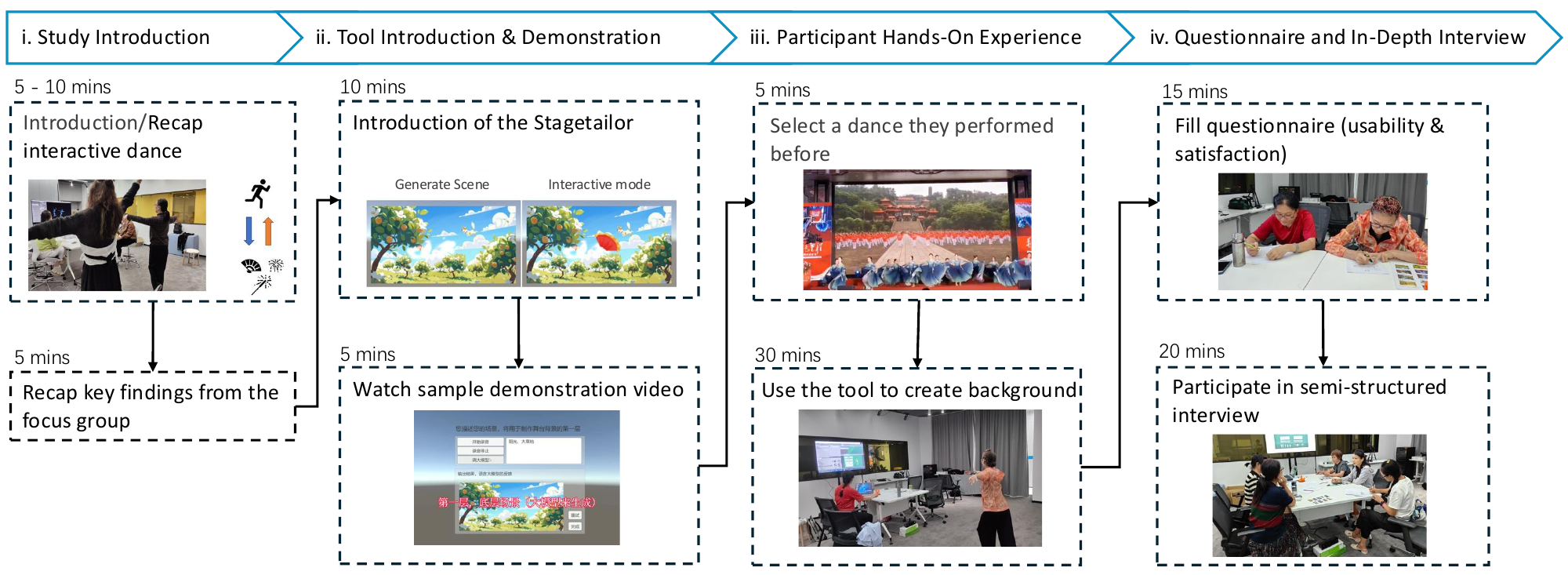}
    \caption{Process of Workshop II with 16 participants divided into four groups (4 participants per group) to evaluate the usability, usefulness, and experiential value of StageTailor. The study followed a four-phase structure: (i) Study Introduction, (ii) Tool Introduction and Demonstration, (iii) Participant Hands-On Experience, and (iv) Questionnaires and In-Depth Interviews.}
    \Description{Process of Workshop II with 16 participants divided into four groups (4 participants per group) to evaluate the usability, usefulness, and experiential value of StageTailor. The study followed a four-phase structure: (i) Study Introduction, (ii) Tool Introduction and Demonstration, (iii) Participant Hands-On Experience, and (iv) Questionnaires and In-Depth Interviews.}
    \label{fig:user study 2 process}
\end{figure*}

\subsubsection{Step 2: Motion Capture and Visual Effects Integration (G2, G3).} 
This step enhances the performance by embedding motion-responsive visual effects that respond to body movements and prop use. This feature enriches the choreography by turning movement into expressive visual storytelling. Technically, we employed a combination of inertial measurement unit (IMU) sensors and visual tracking to capture dancers' motion characteristics, including direction, speed, and gesture dynamics, as well as object-based interactions.

\begin{enumerate}
    \item \textbf{Motion Capture with Multiple Sensors:} Users perform with motion capture devices suited to their setup. These sensors track body acceleration, orientation, and position, triggering corresponding visual effects.

    \item \textbf{Visual Effects Selection and Customization:} Based on the captured motion data, the system recommends thematic effects (e.g., butterflies, ribbons, fireworks). Users can preview, select, and customize them. The final effects are rendered over the background video from Step 1.
\end{enumerate}

We adopted the Kinect system~\cite{kinect} as the primary motion capture device due to its accessibility and reliability. All modules—including motion capture, effect mapping, and rendering—were implemented in Unity~\cite{unity2019}, providing a unified interface for real-time preview and final output.

Together, these components form a cohesive probe that allowed us to investigate
how creative agency, authorship, and embodiment manifest in an \deleted{aging-driendly}\added{age-sensitive} AI-mediated design environment (\autoref{fig:User interface prototype}).

\section{Workshop II: Examining Design Considerations by Co-Creating Stage Backdrops with StageTailor}

With StageTailor established as a probe to examine how the design
considerations (DCs) could be operationalized, Workshop II focused on evaluating
the probe’s usability, usefulness, and experiential value.
Rather than introducing the tool itself, this workshop aimed to understand how
retired dancers engaged with the two-step workflow and how the system shaped
their creative process. The study followed a four-phase format
(\autoref{fig:user study 2 process}).

\subsection{Participants}
A total of 16 retired women (ages 52–67, M = 56.81, SD = 4.20) participated in the study, as shown in Table~\ref{tab:user study participants}. Seven participants returned from Workshop I (P1, P2, P5, P6, P7, P13, P14), providing continuity, while the remaining nine were recruited through snowball sampling from a senior university dance class. All participants had substantial performance experience and regularly engaged in community-based dance activities. To encourage interaction and discussion, participants were divided into four groups of four.

\begin{table*}[htbp]
\centering
\small 
\caption{Participants of workshop II by group. Each group consists of retired women dancers with similar performance histories.}
\Description{This table expands on participant profiles in Workshop II, emphasizing technological familiarity and creative roles. Among the 16 participants (mean age: 56.81), four had prior experience with video editing tools, while two had experimented with AI art generators. The table further identifies emergent roles, such as "Visual Coordinators" (focusing on color and thematic coherence) and "Movement Innovators" (prioritizing motion-triggered effects). Accessibility considerations, such as P14’s reliance on reading glasses and P7’s preference for audio instructions, underscore the need for adaptable interfaces in age-sensitive designs.}
\label{tab:user study participants}
\begin{tabularx}{\textwidth}{@{} p{1.5cm} p{1.5cm} p{1.5cm} p{3cm} p{3.2cm} X @{}}
\toprule
\textbf{Group} & \textbf{ID} & \textbf{Age} & \textbf{Years of Dancing} & \textbf{Freq. of Performances} & \textbf{Team Type} \\
\midrule
\multirow{4}{*}{G1} 
 & P1 & 67 & Over 3 years & Every 2--3 months & Community \\
 & P2 & 60 & Over 3 years & Every 2--3 months & Community \\
 & P3 & 62 & Over 3 years & Every 2--3 months & Senior university / Community \\
 & P4 & 63 & 2--3 years & Every 2--3 months & Community \\
\midrule
\multirow{4}{*}{G2} 
 & P5 & 57 & Over 3 years & Once every 6 months & Senior university / Community \\
 & P6 & 52 & Over 3 years & Every 2--3 months & Senior university / Other \\
 & P7 & 57 & Over 3 years & Every 2--3 months & Senior university / Other \\
 & P8 & 55 & Over 3 years & Every 2--3 months & Senior university / Community \\
\midrule
\multirow{4}{*}{G3} 
 & P9 & 55 & Over 3 years & Every 2--3 months & Senior university / Community \\
 & P10 & 55 & Over 3 years & Every 2--3 months & Senior university / Community \\
 & P11 & 56 & Over 3 years & Every 2--3 months & Senior university / Community \\
 & P12 & 57 & Over 3 years & Every 2--3 months & Senior university / Community \\
\midrule
\multirow{4}{*}{G4} 
 & P13 & 56 & Over 3 years & Every month & Senior university / Community \\
 & P14 & 52 & 2--3 years & Once every 6 months & Senior university \\
 & P15 & 55 & Over 3 years & Every 2--3 months & Senior university / Community \\
 & P16 & 55 & Over 3 years & Every 2--3 months & Senior university \\
\bottomrule
\end{tabularx}
\end{table*}

\subsection{Procedure}

The 100-minute study was conducted in a quiet, well-lit room that provided a distraction-free environment and encouraged active engagement. Each session was facilitated by two researchers and followed a consistent, four-phase process to ensure clarity and comparability across sessions (\autoref{fig:user study 2 process}).

\textbf{Part 1:  Study Introduction.} Participants were introduced to the study's purpose and procedure. To situate the task, they viewed a sample interactive dance video and were reminded of key insights from the Workshop I~\ref{sec:Workshop I}, particularly challenges with static stage backdrops and the potential benefits of dynamic, interactive visuals.

\textbf{Part 2: Probe Introduction and Demonstration.} StageTailor was introduced through a brief presentation explaining its two-step workflow: (1) scene description and AI-generated video creation, and (2) motion-responsive visual effect integration. A live demonstration followed, during which a sample dance video was used to generate a customized interactive background, helping participants understand the system's capabilities.

\textbf{Part 3: Participant Hands-On Experience.} Participants engaged directly with Stagetailor, as shown in~\autoref{fig:User interface prototype}. They were asked to recall a previous dance performance and imagine an ideal visual backdrop. Using either short keywords or full descriptions, they created and refined textual prompts with support from the system. After generating several background options, they selected one to proceed with. In the second step, participants explored motion interaction features. With researcher assistance, they recorded dance movements or prop interactions and selected visual effects from the system’s library. These effects were then integrated into their chosen background, resulting in a personalized performance video. Crucially, researcher assistance was strictly limited to technical support (e.g., system calibration, fault detection, and basic function clarification) and did not involve any guidance on aesthetic choices or creative content, ensuring the autonomy of the participants' creative expression.

\textbf{Part 4: Questionnaires and Semi-Structured Focus Groups.} After experiencing, participants completed a structured questionnaire using a 5-point Likert scale to rate the system's usability, clarity, and perceived value, including their satisfaction with each of the system's key functions: keyword-based description, AI-generated video content, and interactive visual effects. 
This was immediately followed by semi-structured focus groups. The focus groups were conducted with participants grouped by their original dance teams (e.g., Workshop 2 was divided into groups of four), utilizing the inherent community and social dynamics of the retired dancers. This format facilitated a multi-party dialogue where participants reflected on their experience, shared preferences, suggested improvements, and compared the StageTailor approach to traditional methods of background creation. By conducting the focus groups immediately after the creation task, we ensured that feedback was grounded in their fresh, situated experience.

\subsection{Data Analysis}

We employed a mixed-methods approach to analyze data from Workshop 2.

Quantitative Analysis: Data from post-study questionnaires were analyzed by calculating descriptive statistics, including mean scores (M) and standard deviations (SD) across key metrics (e.g., ease of use, perceived usefulness, satisfaction, and perceived creative agency). This evaluated the overall system performance and the magnitude of participants’ self-reported experience.

Qualitative Analysis and Integration: All interview recordings and observations were transcribed and analyzed using thematic analysis with open coding~\cite{corbin2015}. To ensure rigor, two researchers independently coded the data and met iteratively to resolve discrepancies, reaching consensus on the final coding framework. We utilized an explanatory design to integrate the data, where qualitative themes were used to explain and contextualize the numerical quantitative findings. This triangulation approach strengthens the validity of our conclusions regarding creative empowerment.

\subsection{Findings}
\label{subsec:Findings}
This section presents the findings from Workshop II, organized into four key areas: (1) participants' preferences for keyword-based description, (2) their reactions to AI-generated videos, (3) their responses to interactive visual effects, and (4) their perception of the integrated co-creative experience. Overall, the system was perceived by participants as effective in addressing their long-standing challenges with stage backdrops, achieving an average satisfaction rating of 4.07 out of 5—a notable improvement compared to traditional methods.

\subsubsection{Preference for Keyword-Based Description: Lowering Barriers and Facilitating Idea Refinement}

Most participants (14/16) preferred providing keywords rather than full scene descriptions when creating performance backdrops, with an average rating of 3.88 out of 5 for the usability of this feature. Keywords were seen as quicker, less cognitively demanding, and more approachable than writing. Several participants openly admitted discomfort with extended writing (\textit{''I haven’t written anything for years''}, P8), while others valued the immediacy of simply \textit{''throwing in a few words and getting something right away''}. For many, the method made the tool feel accessible rather than intimidating.

\begin{figure}[t]
        \centering
        \includegraphics[width=1.0\linewidth]{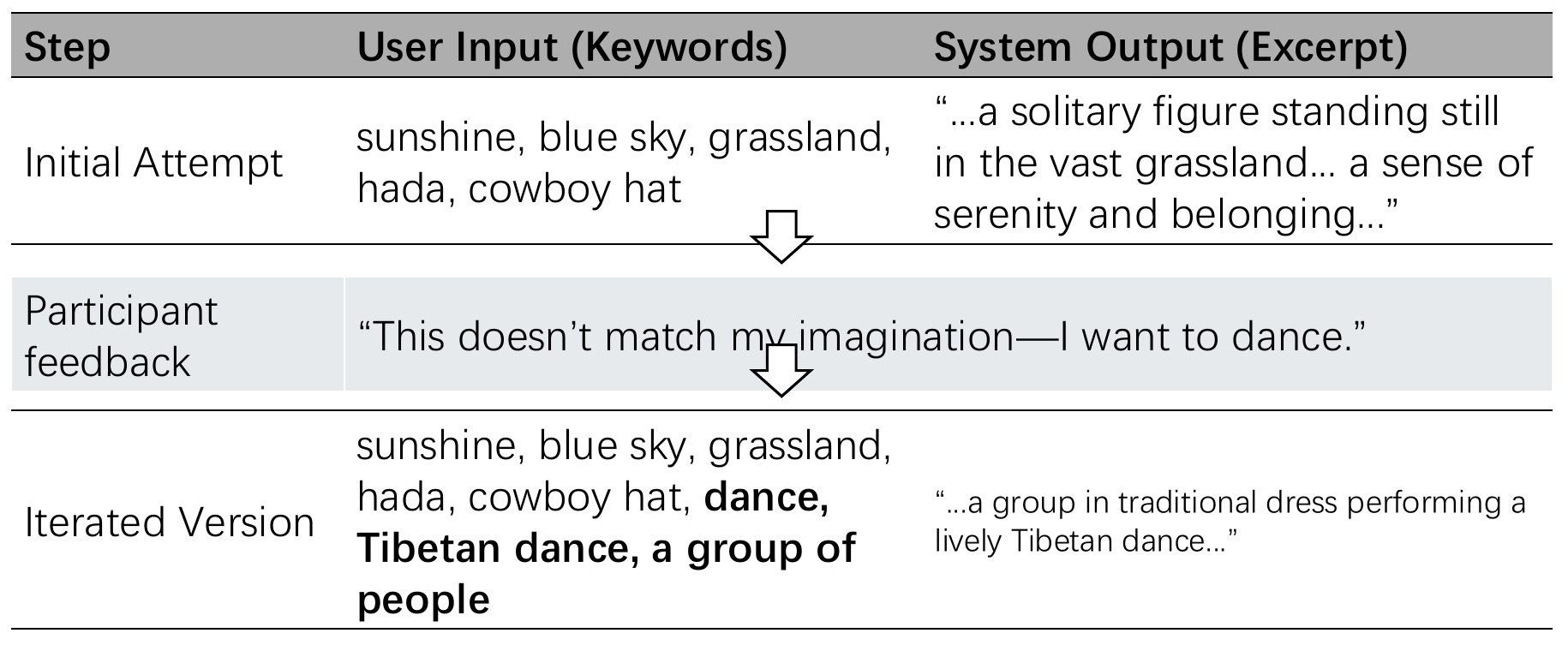}
        \caption{Case Example: Iterative Keyword Selection by One Participant (P10). In her initial attempt, the system generated a solitary figure on a grassland, which did not align with her imagination. After reflecting on the mismatch, she revised her input by adding terms related to dance and groups of people, resulting in an output closer to her envisioned Tibetan dance scene.}
        \Description{Participant P10's iterative design process documented through: (a) Initial sketch: Rough pencil drawing of dancers in circle formation. (b) First AI output: Static landscape with misplaced elements (e.g., cowboy hat in Tibetan scene). (c) Refined version: Dynamic video with 5 dancers in traditional costumes, matching the circular choreography. Callout boxes highlight key changes: Added "group dance" keyword, removed "cowboy hat", adjusted color palette}
        \label{fig:example_keywords}
\end{figure}

Beyond ease of use, the keyword approach actively reshaped how participants engaged with creativity, a sentiment reflected in the high score for creativity stimulation (4.00/5). Instead of requiring them to articulate a fully-formed vision from the outset, keywords allowed them to \textit{''start small''} and progressively refine ideas. Generated descriptions often reminded participants of overlooked details (e.g., adding \textit{''sunrise''} to \textit{''beach''}), or inspired them to expand with symbolic/emotional words such as \textit{''serenity''} or \textit{''energy''}. A typical case was P10 (\autoref{fig:example_keywords}), who initially provided \textit{''sunshine, blue sky, hada (a traditional ceremonial scarf in Tibetan culture), cowboy hat''}, which produced a static landscape. Realizing this did not align with her vision, she refined the input with \textit{''dance''} and \textit{''Tibetan dance''}, resulting in a dynamic multi-performer scene. This illustrates how participants clarified their intentions only after encountering the system’s interpretation --- using the AI not merely as a tool, but as a reflective partner that externalized half-formed thoughts and prompted iteration. In total, participants iterated on their descriptions an average of \textit{2.81 times per person}. This highlights the iterative process of refining creative ideas, with participants adjusting their prompts based on the AI's output and their own evolving vision.

Participants found the keyword-based description method to be exceptionally effective within the context of their retired women's dance group. They (N=7) expressed that a perceived lack of literary and aesthetic expertise made them hesitant to articulate their creative intentions through elaborate narrative or scenographic descriptions. Their creative process had historically been centered on embodied knowledge and movement-based exploration. The keyword approach lowered the barrier to entry, facilitating a fluid transition between their embodied imagination and textual prompts. For example, one participant (P6) conceptually distilled the embodied feeling of 'expansion and release' from a movement into a simple prompt like \textit{'a free-flying bird'}, illustrating the fluid translation from physical to textual expression. This effectively bridged the conceptual gap between their non-verbal choreographic ideas and the demands of digital content creation.

However, participants also reported the limitations. Many participants (11/16) manually revised the AI's outputs, with some finding the language to be too ornate and complex. As P11 stated, \textit{''The wording is so fancy I can’t tell if it’s what I want,''} highlighting a disconnect between the AI’s sophisticated output and their ability to quickly process and evaluate it. This sentiment was reflected in the quantitative data, with the satisfaction score for the generated descriptions averaging a moderate 3.0 out of 5. These findings suggest that while keywords effectively lower barriers and catalyze reflection, future systems must balance creative sophistication with user-centered simplicity. This would empower participants to progressively shape outputs without feeling constrained by an AI's default verbal style or information overload.

\subsubsection{Reactions to AI-Generated Videos: Novelty Meets Expectation Gaps}

\begin{figure}[t]
    \centering
    \includegraphics[width=1.0\linewidth]{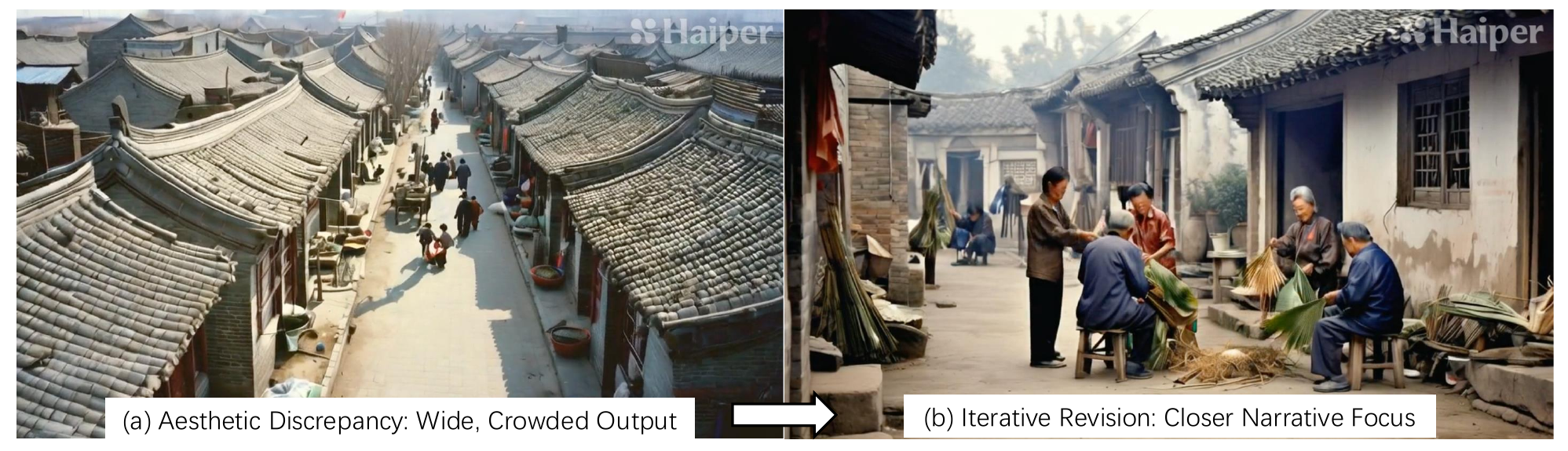}
    \caption{Case Example: P6's Iterative Refinement for an "1980s Chinese Street" Scene. (a) Initial AI output: A wide, crowded street scene that P6 found too busy, obscuring individual details. (b) Refined AI output: A closer view with more distinct characters, achieved after P6's iterative prompt adjustments, better aligning with her desire to "see each person's role."}
    \Description{A two-part image showing AI-generated video stills. (a) A wide shot of a bustling 1980s Chinese street with many people, appearing somewhat generic. (b) A closer, more focused shot of a few distinct individuals on a similar street, showing more detail and character.}
    \label{fig:p6_example}
\end{figure}

Participants responded to the AI-generated videos with initial excitement but mixed long-term satisfaction (average rating 3.33/5 for the satisfaction with the generated video content). For many (15/16), the mere act of watching their words turn into animated scenes was novel and inspiring. Exclamations like \textit{''How amazing!''} and \textit{''This is really interesting!''} reflected a sense of wonder, particularly among those unfamiliar with generative media.

Yet this novelty quickly gave way to critical reflection when participants compared outputs with their envisioned stage settings, especially regarding the finer details and subtleties of the scene. Twelve participants wanted to revise or regenerate videos, often due to mismatches in atmosphere, style, or detail. For example, P9 described a Tang or Song dynasty school at sunrise, but the generated video lacked architectural cues and appropriate lighting. Others noted issues with composition, wide, crowded shots that obscured characters (P11, P6), or missing temporal flow for story-like scenes. These gaps revealed a deeper challenge: participants struggled to translate nuanced, embodied visions (lighting, pacing, emotion) into text prompts that current systems could interpret. On average, participants iterated on their videos \textit{3.25 times per group}, indicating the iterative nature of refining the output. Many wanted to refine the generated content to better match their envisioned settings.

This process, however, also revealed how participants shifted from being passive recipients of visuals to active critics of digital outputs. They identified what was “off” and suggested improvements such as using reference images, storyboards, or multi-stage generation to capture unfolding narratives. For instance, P6 envisioned an everyday 1980s Chinese street with people of different ages, but the output collapsed this into a single wide-angle frame (\autoref{fig:p6_example}a). Her frustration (\textit{''It’s too crowded; I wanted to see each person’s role''}) pointed to an unmet need for narrative scaffolds, not just single-scene outputs. Through iteration, P6 managed to refine her prompt to generate a closer view (\autoref{fig:p6_example}b) that better highlighted individual characters, though still not fully capturing the multi-generational narrative she desired.

Despite these limitations, participants unanimously agreed that generating videos on demand was far superior to searching for stock footage or settling for static backgrounds. As P11 summarized: \textit{''Now we can generate scenes that match our ideas instead of endlessly searching and compromising.''} This balance—between amazement, critique, and constructive suggestion—shows that generative video can empower older dancers as creators, but only if future systems support iterative refinement and provide more visual, narrative-oriented modes of control.

\subsubsection{Responses to Interactive Visual Effects and System Feedback: Engagement but Limited}

Participants responded enthusiastically to the interactive visual effects, with an average rating of 4.0 out of 5 for their satisfaction with the interactivity of the visual effects. Thirteen participants (81.3\%) emphasized how the responsiveness transformed their performance experience: \textit{''Unlike the rigid, unrelated backgrounds of the past, these respond to what I’m doing''} (P16). This novelty created a sense of co-creation, with several dancers describing the feeling of “shaping the entire visual experience” (P1).

At a deeper level, these effects reframed participants’ relationship with their own movement. Instead of focusing on precision, dancers began to view their gestures as triggers for visual storytelling. For older adults who sometimes struggled with synchronization or technical execution, this reorientation provided confidence: even small movements could produce meaningful effects. As one dancer noted, \textit{''The butterfly followed my hand—I felt like my dance was alive.''}

Yet, limitations quickly surfaced. The average satisfaction rating dropped from 4.0/5 for the interactive preview to 3.0/5 when the final effects were composited onto videos. This drop in satisfaction was not due to a simple technical fault, but a conceptual challenge in our prototype’s design: the inherent tension between real-time interactivity and a pre-generated video background. Participants criticized the lack of fluidity and abrupt transitions between effects and backgrounds. For instance, a participant (P7) who chose a butterfly effect on a grassland background noted how the butterfly's flight path would unnaturally cut through the animals and trees in the scene. Another participant (P6) questioned if a motion-aligned “sheep” effect would simply fly across the screen like a butterfly, highlighting the semantic mismatch between the visual effect and its background. These issues reduced the sense of immersion and coherence, leaving some to describe the feature as promising but not yet performance-ready.

Still, the feature revealed an important design opportunity: interactive visuals can shift focus from technical execution toward expressive engagement, making performances feel richer even in resource-constrained settings. What participants sought was not simply “more effects,” but smoother integration, thematic variety, and expressive control that could accommodate diverse dance styles.

\subsubsection{Integrated Experience: From Users to Co-Creators}

Participants’ reflections on the system went beyond utility or convenience; many emphasized a shift in their role. Instead of being passive performers who danced in front of borrowed or pre-made visuals, they felt like co-creators of the stage itself. One participant summarized this transformation: \textit{''This tool finally lets me create something that feels like mine—not just something copied from online''} (P6).

This sense of ownership emerged through the interplay of AIGC and interactive visuals. By providing input—even as simple as a few keywords—participants saw their ideas materialize on stage. This process fostered empowerment, which was reflected in their high willingness to participate in stage creation (3.88/5). This transformation gave retired women not only tools for performance but also a voice in shaping their artistic environment. Several explicitly contrasted this with past experiences of “waiting for someone else to decide the background” (P4), noting how the system made them feel part of the creative team.

At the same time, the system fostered agency and meaning. Participants described the outputs as \textit{''more personal''} (P13) and \textit{''like performing somewhere professional, even with just our community’s single screen''} (P2). The ability to integrate personalized visuals—despite limited resources—made their performances feel distinctive and resonant, reinforcing the idea that technology could bridge the gap between amateur community dance and professional stage production.

Together, these insights demonstrate that accessible AIGC systems can fundamentally redefine the user’s role in community dance: transitioning them from passive consumers of borrowed media into empowered co-creators who actively imbue their performances with personal and collective meaning.

\section{Discussion}
The results of our Workshop I identified challenges for retired women dancers in enhancing the expressiveness of their community performances, especially dissatisfaction with irrelevant, repetitive, and hard-to-customize backdrops due to limited resources, skills, and time. To address this, we developed StageTailor, integrating LLM-powered scene description, AI-generated video, and interactive visual effects, as a creative probe in a co-design study with the retired women dancers. Findings suggest this tech-supported tool improved efficiency, sparked creativity, and shifted participants from passive users to active co-creators, yet there exist several challenges. These included a mismatch between the AI's language style and the participants' cognitive load, the difficulty in translating multi-layered creative intent into effective text prompts, and a conceptual mismatch between interactive effects and pre-generated video backgrounds. Based on these findings, we propose eight Design Implications (DIs) that address the specific needs of this group while contributing to broader HCI discussions on aging, creativity, and AI-supported design.

\subsection{Rethinking Keyword Input as a Reflective Interaction Process}

Our study shows that keyword-based input was not only more accessible for older adults than full textual descriptions, but also more effective in stimulating their imagination. Participants, who often described themselves as "out of practice with writing" or "never good at language", preferred to enter simple, symbolic words. Importantly, these inputs did not simply generate results; they served as low-barrier entry points for reflection. The pattern of “input – reflection – adjustment” highlights the value of treating keyword input not as a static command, but as a dynamic and co-creative dialogue with the system, helping participants clarify their unspoken ideas.

This resonates with recent HCI work calling for more playful and expressive interaction modes for older adults in creative AI contexts~\cite{qin2024charactermeet}. Prior studies in intergenerational co-creation have similarly emphasized the power of combining minimal verbal input with visual scaffolds~\cite{kim2025bridging}. Our findings extend these insights to non-professional AIGC prompt design, showing that even vague expressions can initiate rich creative cycles when supported by iterative, visually grounded feedback.

\subsubsection{DI1: Support Reflective and Guided Keyword Input.}
For this group—less accustomed to writing but comfortable with embodied expression—systems should embed lightweight guidance such as category prompts or example pools. This is especially valuable for older adults, who are less accustomed to abstract prompt-writing, but can build on familiar categories and concrete examples. Indeed, prior work on ideation tools for these populations has demonstrated that visual prompts and structured input mechanisms can reduce decision fatigue and boost user confidence~\cite{10.31234/osf.io/zd82b,10.1145/2757226.2764773}. For example, a system could provide thematic buttons like Forest, Ocean, and City, which then populate a sub-menu with more specific, contextually relevant suggestions like Serene Pond or Energetic Waterfall.

\subsubsection{DI2: Provide Visual Selectors for Visual Attributes.}
Participants struggled to articulate visual features such as lighting, perspective, or atmosphere. Given retired women dancers’ reliance on visual and embodied thinking over textual articulation, systems should provide galleries of visual attributes, enabling intuitive expression without requiring technical terminology. This difficulty resonates with findings in creative co-authoring and video storytelling, where visual references help bridge vocabulary gaps~\cite{10.1017/pds.2021.439,10.1145/317561.317590}. For example, an interface could display a gallery of lighting presets (e.g., thumbnails showing Golden Hour, Moonlight, Spotlight) or camera angles (e.g., a series of images depicting Wide Shot, Close-up, Low Angle) that users can simply select to apply their desired effect.

\subsection{Bridging the Gap Between User Intent and AI Video Output}
Although participants found the AI-generated background videos novel and engaging, many reported that the final outputs did not fully reflect their envisioned scenes. As retired dancers preparing content that complements choreography, participants often had vivid mental imagery but lacked the vocabulary to express it precisely. This created a challenge where they expressed uncertainty about how to improve a scene, sometimes choosing to accept the result by default.

These challenges echo findings from AI co-creation tools, where users struggle to align symbolic or emotional intentions with system-generated media due to limited familiarity with visual storytelling conventions or prompt engineering~\cite{kim2025bridging, li2024flowgpt}. Our work highlights this gap specifically in the context of visual storytelling for performance, where spatial pacing, rhythm, and mood are especially difficult to convey textually.

\subsubsection{DI3: Offer Multiple Differentiated Outputs for Reflective Selection.}
When outputs were misaligned with their vision, participants often accepted results due to difficulty articulating changes. However, presenting alternatives helped them recognize preferences, echoing prior work on example-based refinement showing that branching outputs reduce reliance on precise verbal commands~\cite{li2024flowgpt}. For this population, multi-option outputs can scaffold reflection, supporting iterative alignment without technical burden. For example, after a user provides a prompt like "sunny mountain scene," the system could present three distinct versions: a photo-realistic mountain with a clear sky, a painterly rendition with soft clouds, and an animated scene with sun rays filtering through trees. By having these options, users can simply click the one that feels closest to their vision and refine from there.

\subsubsection{DI4: Scaffold Storytelling with Emotion- and Time-Based Structures.}
Participants frequently imagined visuals that matched emotional progression or choreography rhythm but lacked vocabulary to describe visual arcs. Building such structures into AI video generation is especially important for retired women dancers, whose expression often centers on symbolic or emotional cues rather than technical film language. Creativity support literature highlights the value of scaffolds that translate affective goals into structured input~\cite{10.1145/3290605.3300619,10.1145/1978942.1979048}. For instance, a system could provide a simple timeline interface where users can select points in a dance and assign emotional tags like "calm," "excitement," or "sorrow." The AI could then generate a video background that shifts from a serene forest to a stormy sky to match the emotional arc of the performance.

\subsection{From Reactive Effects to Context-Aware Visual Integration}
Participants responded enthusiastically to the system’s interactive visual feedback feature, especially the way visual effects responded in real-time to their dance movements. This dynamic responsiveness contrasted sharply with the static or pre-recorded backgrounds they had previously used, creating a sense of agency and co-creation. Prior work in performance-based interaction emphasizes the importance of coherence between bodily gestures and visual feedback~\cite{10.1038/srep17657}, while studies in affective computing show that aligning visuals with emotional tone enhances immersion and user satisfaction~\cite{kim2025bridging}. Despite this excitement, participants also surfaced limitations, describing the effects as “superimposed,” lacking depth or integration with the background.

\subsubsection{DI5: Map Movement Qualities to Genre-Aligned Visual Feedback.}
While participants appreciated responsive effects, they critiqued them as repetitive or out of sync with the performance. The issue was not a lack of immediate synchronization, but rather a conceptual gap between the system's simple responsiveness and the expressive fluidity of the dance. For this group, mapping motion features (tempo, intensity, fluidity) to genre-appropriate aesthetics (e.g., gentle glows for waltz, bursts for folk dance) can amplify both bodily expression and cultural resonance. Prior work on embodied interaction and gesture-based performance demonstrates that aligning gesture qualities with expressive visuals enhances immersion~\cite{10.20944/preprints201810.0527.v1,10.31234/osf.io/qfc9e}. For example, for a traditional Chinese fan dance, the system could detect a fluid, graceful arm movement and respond with elegant, flowing silk ribbons that visually extend the dancer’s gesture. Conversely, for a more energetic folk dance, a sharp, percussive foot stomp could trigger a burst of vibrant, confetti-like particles, directly amplifying the rhythm and intensity of the movement. This genre-aligned mapping transforms simple responsiveness into meaningful, culturally resonant co-creation.

\subsubsection{DI6: Ground Interactive Feedback in Scene Semantics.}
Participants described effects as “pasted on,” disconnected from the AI-generated backdrops. Designing feedback that draws materials from the generated background (e.g., leaves responding to movement in a forest scene) can create richer, more meaningful integration for community performances. Scene-aware augmentation research shows that grounding visuals in the narrative environment strengthens coherence~\cite{10.1002/mar.21632,10.3389/frobt.2018.00037}. For example, if the background is a grassland, the interactive effect could be a motion-responsive rippling of the grass itself or a flock of birds scattering in response to a dancer's gesture. Instead of a butterfly rigidly following a dancer's hand, it could be designed to flutter around the dancer, occasionally landing on a flower or a blade of grass, creating the illusion of a collaborative dance between the visual effect and the performer.

\subsection{Reframing Older Adults as Empowered Co-Creators}

Participants appreciated not just the creative output but the sense of authorship and agency that StageTailor enabled. Many described feeling more “involved,” “creative,” and “in control” than in past performances. Our study extends these principles into the domain of stage production—where creative roles are often inaccessible to non-professionals. Crucially, we argue that this shift from performer to co-creator is facilitated by a novel approach to AI mediation, which can be conceptually distinguished from prior work on older adults’ digital creativity.

Previous HCI work has established that older adults can serve as competent content creators, particularly in domains focused on personal narratives, social connection, and digital storytelling~\cite{ferreira2016learning, lazar2016designing,waycott2013older}. This capability is vividly demonstrated in their widespread participation on Short-Form Video Sharing Platforms (SVSPs), where low technical barriers and algorithmic affordances empower them to transition from lurkers to active contributors, driven by the enjoyment and social support they receive~\cite{tang2023towards}. And recent work actively employs AIGC/LLM to lower barriers and expand creative possibilities. For instance, studies have explored using LLM-enhanced robots that act as a "coach" in collaborative painting, providing inspiration through dialogue~\cite{bossema2025llm}. Similarly, systems like JournalAIde demonstrate how LLMs can scaffold the digital diary writing process by helping older adults brainstorm ideas and refine their personal narratives~\cite{zhou2025journalaide}.

While reinforcing this capacity, StageTailor’s contribution lies in its distinct domain and mechanism. These prior studies largely examine content creation in everyday, low-stakes contexts (e.g., photography, digital stories). Our work extends this scholarship into a domain largely overlooked: community-based, high-visibility stage performance, where visual design roles are traditionally inaccessible to non-professionals. This shift from documenting life to mastering stagecraft introduces distinct challenges—coordination with choreography and high aesthetic expectations—requiring new forms of technological mediation. The novelty is in how an AI system can successfully bridge the gap between low-barrier input and complex, professional-level live artistic output, moving beyond feasibility to demonstrate a specific conceptual contribution to aging HCI. \added{This also suggests that designing for age-sensitive creative AI should focus less on eliminating competence or expertise, and more on mediating between users’ existing experiential knowledge and its computational representation. In our case, participants’ embodied understanding of movement, rhythm, atmosphere, and stage aesthetics did not need to be replaced or simplified, but translated into forms that AI systems could meaningfully interpret and extend.}

\subsubsection{DI7: Embed Visible Attribution to Strengthen Creative Ownership.}
Participants expressed pride in seeing their ideas visibly shape the performance, with one stating this was the first time a backdrop “felt like it belonged to me.” To foster this sense of ownership, designs should employ strategies to make users' contributions tangible and traceable. Visible attribution is a powerful method for this group, as it transforms their internal sense of authorship into an external, affirmed reality. Research on older adults’ creativity highlights that authorship and visible recognition drive sustained engagement~\cite{10.1017/s0144686x20000495,10.1111/bjdp.12043}. For instance, a future system could dynamically display a contextual label on the video or interface, such as "Scene inspired by your keyword 'serene forest'," or "Lighting by P7." When a created scene is shared within a community library, the original creator's name or username could be prominently displayed to reinforce their contribution.

\subsubsection{DI8: Support Collaborative and Community-Based Co-Creation.}
Although our study focused on individuals or pairs, participants expressed interest in group design, echoing findings from intergenerational co-creation platforms that emphasize shared agency~\cite{kim2025bridging}. Supporting collaborative scene-building or mentoring modes could resonate strongly with retired women dancers, whose practices are deeply embedded in social and community bonds. This design implication opens several promising avenues for future research, including:

\textbf{Collaborative Prompt Engineering:} Future studies could explore how to design interfaces that allow multiple participants to collectively contribute keywords or edit videos simultaneously. For example, a shared canvas mode could let dancers work on the same scene in real time, or role-based collaboration could be explored, where one user focuses on the background scene and another on the interactive effects.

\textbf{Community-Driven Content Libraries:} Researchers can investigate how to build a community-driven content library where participants can share their generated videos, effects, or creative templates. This would allow for a "remix" culture to emerge, empowering the community to collectively build and curate their own library of performance assets.

\textbf{Mentorship and Skill Scaffolding:} Future work could explore how a system can facilitate mentorship between more experienced users (e.g., those with a background in art or technology) and newer members. This would not only enhance skill-building but also strengthen community ties and provide a more personalized creative support system.

By designing and studying StageTailor as a research probe, we propose an emerging design direction for HCI: \deleted{aging-friendly}\added{age-sensitive} creative AI mediation. These systems use AI not as a replacement for creativity but as a mediator that helps older adults move from performers to co-creators of complex, multimodal artistic experiences. This mediation model prioritizes aesthetic control and low-cognitive load input over technical complexity. This enables older users to directly influence core aspects of their performance environment—scenography and motion-reactive effects—a creative power previously gated by technical expertise.

This reframing aligns with and extends several theoretical traditions in HCI. Specifically, it relates to design-for-agency scholarship~\cite{vines2015age}, which emphasizes supporting older adults’ capacity to act, influence, and see the consequences of their actions in the world. Furthermore, it resonates with the core ethos of Participatory Design (PD)~\cite{bodker2018participatory,bannon2018introduction}, which advocates for the redistribution of decision-making power to users often marginalized in technology design processes. While our work was not a full-scale PD engagement (i.e., it did not involve deep, longitudinal co-creation across the entire design cycle), we describe the work as having a "participatory flavor" because its central goal was the design of "participatory scaffolds." These scaffolds (low-barrier prompting, motion-aligned control) operationalize the PD principle by making creative decisions accessible.

StageTailor operationalizes this by grounding its design in the COM-B framework principles~\cite{west2020brief}: it expands participants’ capability (by scaffolding technical expression through AIGC/LLMs), opportunity (by granting access to the stage design domain), and volition (by strengthening creative ownership through immediate and personalized feedback). By providing \deleted{aging-friendly}\added{age-sensitive} scaffolds, the system effectively redistributes creative control in a typically professional domain, enabling older women to participate meaningfully in decisions about scenography and visual expression, thus cultivating empowered creative subjectivity in later life.

\subsection{Critical Reflections on Creative Agency and AI Mediation}
Facilitating the workshops with retired dancers deepened our understanding of creative aging. Although we initially focused on the technical mechanisms that support creativity, it became clear that creative agency is equally rooted in perception, ownership, and self-efficacy. Participants described feeling more “involved,” “creative,” and “in control” than in past performances, highlighting how psychosocial factors shape their engagement with creative tools in later life.

Our findings suggest that AI can mediate creativity without replacing human intention. StageTailor amplified participants’ situated and embodied inputs (keywords, gestures, and movements) into complex artistic outcomes, allowing them to retain editorial and aesthetic authority. This resonates with co-creation perspectives in HCI, positioning AI as a creative partner that translates user intent rather than generating content autonomously.

At the same time, our study highlights a tension in AI-assisted creativity. AIGC tools make artistic production more accessible, but they also complicate the question of what counts as a “creative act.” Simply generating an image or video through AI does not automatically clarify who the “author” is or what creative labor means in this context. However, our findings show that participants’ sense of creativity did not come from technically producing the pixels, but from the decisions they made throughout the process, choosing keywords, adjusting scenes, selecting effects, and interpreting the results. Their feeling of authorship emerged from this ongoing process of directing and shaping the AI’s output. This suggests that in AIGC settings, creative agency is not located in the final artifact but in the interaction itself: how people guide, critique, and refine what the system produces. In this sense, creativity becomes a relational process between human and machine, rather than something defined by technical authorship alone.

\section{Limitations and Future Work}

\subsection{Limitations}
While our study offers valuable insights into the design of AI-supported creative tools for retired women dancers, it has several limitations that provide clear avenues for future research. First, our sample was specific to a single community in China with prior performance experience. This limits the generalizability of our findings to other groups, such as those with different cultural backgrounds, genders, or varying levels of digital literacy. Second, our study's short-term, workshop-based nature means we cannot speak to the long-term usability and creative impact of the system in a real-world performance setting. Third, our study did not address the economic costs associated with using AIGC models. As AIGC tools introduce new expenses, the question of long-term economic viability is crucial for real-world deployment. Future research must systematically investigate the cost-effectiveness of tools like StageTailor.

\subsection{Future Research Opportunities}
Beyond the immediate findings, this work opens a broader space for exploring \deleted{aging-friendly}\added{age-sensitive} creative AI mediation—systems in which AI serves as a scaffold, amplifier, and collaborator in older adults’ creative expression. We outline four directions that extend both our design implications and the conceptual contribution of this study.

\textbf{Optimizing Creative Language and User Cognition}
Our study revealed that the complexity of AI-generated content can sometimes exceed users' cognitive load, leading to a mismatch in style. Future research can explore how to design more intelligent systems that engage in a more natural and user-friendly creative dialogue. For instance, systems could explore developing language models that can automatically adjust the complexity and style of the generated content based on a user’s input style, such as symbolic, emotional, or descriptive.

\textbf{Bridging the Intent-Output Gap}
We found that participants struggled to translate complex, multi-layered creative intents—such as composition, rhythm, and emotion—into precise text prompts. Future work should focus on helping users better control and shape the AI's output. For example, systems could explore a visually-driven creative process that first generates a range of stylized videos with different compositions, lighting, and colors, allowing users to select and adjust by clicking rather than relying solely on text-based modification.

\textbf{Enhancing the Integration and Co-creation of Interactive Effects}
A significant challenge we identified was the conceptual mismatch between interactive effects and static backgrounds. Future work needs to move beyond simple “motion-following” functionalities to achieve a deeper level of visual fusion and co-creation. For instance, researchers can investigate how to develop AI models that can understand the semantics of a scene. When a user dances in a "forest" scene, the system could intelligently generate effects like rustling leaves, making the effect feel like an organic part of the scene rather than a simple overlay.

\textbf{Fostering Community and Collaborative Creation}
Our study revealed the importance of community and collaboration for this group, which directly aligns with our DI8. Future work should aim to build an ecosystem that supports collective creation and sharing. For example, researchers could explore how to design a platform that allows multiple users to collaborate on a single scene, or establish a community-driven library of backgrounds, effects, and templates that users can co-create and maintain. This would effectively leverage the collective intelligence of the community while further enhancing their sense of authorship and belonging.

Together, these directions suggest that the core contribution of this work, \deleted{aging-friendly}\added{age-sensitive} creative AI mediation, can be expanded far beyond stage performance. Future applications may include community filmmaking, AR-based participatory exhibitions, intergenerational craft design, or AI-supported cultural preservation. Across these domains, the central challenge remains the same: designing AI systems that enhance older adults’ capability, opportunity, and volition to act as empowered creative agents in later life.

\section{Conclusion}

This paper explored how interactive dance and AI-powered video generation can support retired women dancers in their community performances. Through two co-design workshops with 16 participants, we demonstrated that \deleted{aging-friendly}\added{age-sensitive} adaptations, such as low-barrier keyword input, motion-aligned effects, and participatory design scaffolds, successfully lower technical barriers and foster creativity. These features empowered participants to transition from passive consumers of pre-made visuals to active co-creators of their stage environments. Our work contributes to aging-and-creativity research by illustrating how older adults can engage in AI-supported artistic expression, and offers broader strategies for designing accessible creative technologies.



\begin{acks}
\small
This research is partially funded by 2025 Guangdong Undergraduate University Teaching Quality and Teaching Reform Project, Red Bird MPhil Program, AI Research and Learning Base of Urban Culture under Project 2023WZJD008, Guangdong Provincial Key Lab of Integrated Communication, Sensing and Computation for Ubiquitous Internet of Things (No. 2023B1212010007), and the Project of DEGP (No. 2023KCXTD042).

We used Large Language Models (LLMs) such as ChatGPT-4, Gemini 2.5, and DeepSeek as translation aids to translate our original manuscript from Chinese to English. The translated text was thoroughly reviewed and edited by the authors to ensure accuracy and academic integrity. The authors take full responsibility for the content and use of AI in this paper. In the development of our research probes, we also leveraged AI tools, specifically Ernie Bot for Chinese-language LLM support and Haiper for text-to-video generation.

We also wish to thank the anonymous reviewers for their invaluable feedback and suggestions, which significantly improved the quality of this paper.
\end{acks}

\bibliographystyle{ACM-Reference-Format}
\bibliography{main}

\end{document}